\documentclass[useAMS,usenatbib]{mn2e}
\usepackage{epsfig}
\usepackage{amsmath}
\begin{document}

\title[The asymmetric structure of the Galactic halo]{The asymmetric structure of the Galactic halo}

\author[Y. Xu, L. C. Deng and J. Y. Hu]{Y. Xu$^{1,2}$\thanks{E-mail:
xuyan@bao.ac.cn(YX); licai@bao.ac.cn(LCD); hjy@bao.ac.cn(JYH)}, L.
C.
Deng$^{1}$\footnotemark[1] and J. Y. Hu$^{1}$\footnotemark[1] \\
$^{1}$National Astronomical Observatories, Chinese Academy of
Sciences, Beijing 100012, P. R. China\\
$^{2}$Graduate University of Chinese Academy of Sciences, Beijing,
100049, P. R. China}
\pagerange{\pageref{firstpage}--\pageref{lastpage}} \pubyear{2002}

\maketitle

\label{firstpage}

\begin{abstract}
Using the stellar photometry catalogue based on the latest data
release (DR4) of the Sloan Digital Sky Survey (SDSS), a study of the
Galactic structure using star counts is carried out for selected areas
of the sky. The sample areas are selected along a circle at a Galactic
latitude of +60$^\circ$, and 10 strips of high Galactic latitude along
different longitudes. Direct statistics of the data show that the
surface densities of $\ell$ from $180^{\circ}$ to $360^{\circ}$ are
systematically higher than those of $\ell$ from $0^{\circ}$ to
$180^{\circ}$, defining a region of overdensity (in the direction of
Virgo) and another one of underdensity (in the direction of Ursa
Major) with respect to an axisymmetric model. It is shown by comparing
the results from star counts in the $(g-r)$ colour that the density
deviations are due to an asymmetry of the stellar density in the
halo. Theoretical models for the surface density profile are built and
star counts are performed using a triaxial halo of which the
parameters are constrained by observational data. Two possible reasons
for the asymmetric structure are discussed.
\end{abstract}


\section{Introduction}

It has been one of the central goals of human knowledge to understand
the position of the Earth in the universe. A lot of effort has been
spent to derive the structure and the history of formation and
evolution of our Galaxy. Doing this is a particularly difficult task,
simply because there is no good viewpoint, as opposed to the case for
external galaxies whose structures can be observed directly. The
structure, formation and evolution of the Galaxy are very important
issues in contemporary astrophysics; they are closely related to
stellar quantities such as distance, age, metallicity, and kinematics
of the individual stars, among others. These properties can be
obtained for large stellar samples using straightforward photometric
and spectroscopic observations. In the present era, before the full
exploitation of the planned huge spectral surveys of Galactic stellar
objects (e.g. GAIA, SEGUE, 6dF, LAMOST) is possible, using star counts
based on all-sky photometric surveys is a direct and one of the few
accessible methods for the study of the structure of the Galaxy.

Probing the Galaxy using star counts has a history of about 200 years
(Reid 1993), with the modern epoch marked by the classic work of
Bahcall \& Soneira (1980), based on several observations (Seares et
al. 1925, Kron 1978, Tyson \& Jarvis 1979, Peterson, Ellis \&
Kibblewhite 1979). Bahcall \& Soneira built a standard model, in which
the Galaxy was simplified and parameterized by an exponential disk and
a spheroid characterized by a de Vaucouleurs profile. Assuming a
density profile for each population, based on detailed studies of
nearby spiral galaxies, and assuming a luminosity function as observed
for stars in the local volume, their model predicts number density --
magnitude -- colour distributions.

Further studies on this subject provided certain modifications to
the Bahcall \& Soneira model. Theoretical models were further
constrained using new observations were subsequently obtained. In
1984, Gilmore claimed for the first time that there is a thick
disk in our Galaxy. The star counts approach to the study of the
structure of the Galaxy was extensively reviewed by Reid (1993);
he, and his colleagues, (Reid 1993, Reid, Yan \& Majewski 1996,
Majewski, Siegel \& Kunkel et al. 1999, Siegel, Majewski \& Reid
2002) carried out a series of studies appropriately termed ``star
counts redivivus'', dedicated respectively to: faint magnitude
star counts, the halo luminosity function, and an exploration of
the contamination of star counts by star streams. When the Sloan
Digital Sky Survey (SDSS) made its first data release publicly
available, it raised very promising prospects for the study of
Galactic structure thanks to the large sky coverage and the deep
stellar photometry. Chen et al. (2001) analysed the SDSS early
data release (EDR); they confirmed the existence of a Galactic
thick disk and obtained a set of structural parameters for the
Galaxy. With the increasing sky coverage in subsequent SDSS data
releases, it is expected that one will be able to determine the
Galactic structure parameters ever more precisely. Except for
constraining the density profile, the stellar luminosity function
can also be determined with a higher accuracy. In Gould, Bahcall
\& Flynnet (1996), and Zheng, Flynn \& Gould (2001, 2004) the disk
M-dwarf and $I$-band luminosity functions are obtained using {\sl
Hubble Space Telescope} star counts. In Reid (2002), the
luminosity function in the solar neighborhood was studied based on
the Palomar/MSU nearby star spectroscopic survey.

The Galactic structural parameters derived from various studies show
large divergence, which cannot simply be attributed to statistical
errors, although these are generally significant. In star counts from
fields selected following a certain symmetry approach, there are
deviations from the axisymmetric standard model which are obvious and
cannot be neglected. Spergel \& Blitz (1988) pointed out that this
might be an indication of a triaxial structure of the Galaxy, and they
used a triaxial halo to explain the asymmetries detected from gas
motions around $\ell$=180$^\circ$. Later, Blitz \& Spergel (1991)
showed that the observed H{\sc i} gas is moving in an asymmetric
gravitational potential. Using areas selected from the ASP catalogue,
Larsen \& Humphreys (1996) found deviations of star counts from the
axisymmetric model; they found that there are significant excesses of
stars in quadrant I, at $l=20^{\circ} - 50^{\circ}$, compared to
quadrant IV, around $b=30^\circ$. Further work was published by
Parker, Humphreys \& Larsen (2003), who compare star counts in 120
POSS I fields. They claimed that the excess is not associated with any
known stream such as those linked to the Sagittarius dwarf, Magellanic
stream or Fornax-Leo-Sculptor stream. Parker, Humphreys \& Beers
(2004), however, concluded that the excess is due to thick disk stars
and may be associated with the recently discovered Canis Major debris
stream. In Hartwick (2000), possible reasons for causing triaxiality
due to nearby spiral and dwarf galaxies are considered. Newberg \&
Yanny (2005b) built triaxial ellipsoidal stellar halo models and
analysed F turn-off star samples from SDSS DR3.

To explain the observed excess of the stellar number density with
respect to an axisymmetric halo, two competing scenarios have been
proposed: the triaxialilty could be an inherent property related to
the smooth structure of the Galaxy, or it could be related to the
remnants of some historical merger events. Recently, this debate has
intensified with the reported ``Virgo overdensity'' by Juri\'c et
al.(2005), which may be the cause of the large-scale asymmetry in the
structure of the Galaxy. By using the photometric parallax method,
they measured stellar distances which allowed them to construct a map
of the 3D structure of the sky area covered by current SDSS survey.
Their study has, for the first time, revealed large-scale patterns in
the Galaxy. In the same data that showed the Virgo overdensity, an
underdense area with respect to an axisymmetric halo model is also
obvious in the direction of UMa. Whereas the overdensity can be
naturally interpreted as a large-scale star stream in the halo model,
the UMa underdensity cannot be accounted for in the same way. We
believe that this observational fact is linked to the smooth structure
of the halo in which the star streams (including the Virgo
overdensity) are embedded. To tackle this problem, a similar star
count analysis of Galactic structure will be followed in this paper.

The Sloan Digital Sky Survey is a five-colour ($u, g, r, i, z$)
broad-band photometric survey with a wavelength coverage from 3000 to
11,000{\AA} (Fukugita, Ichikawa \& Gunn 1996). The DR4 imaging
catalogue covers 6670 square degrees. Its detection repeatability is
complete at a 95\% level for point sources brighter than the limiting
apparent magnitudes of 22.0, 22.2, 22.2, 21.3 and 20.5 for $u, g, r,
i$ and $z$, respectively (http://www.sdss.org).  It has facilitated
great progress in the study of the structure of the Galaxy, which was
reviewed in the SDSS-II SEGUE Project. For the study of Galactic
structure using star counts, SDSS provides the most up-to-date and
complete stellar sample in terms of both sky coverage and magnitude
limit. Thus, the SDSS is suitable to study the large-scale asymmetry
of the Galaxy. In this paper, we build a model of the Galaxy with a
non-axisymmetric geometry, and constrain the parameters of the model
using observations in a set of characteristic directions taken from
SDSS DR4. In Section 2, the selected observational data are
described. In Section 3, the model used to fit the observational data
is introduced. The results of triaxial halo model and the asymmetric
structure of the Galaxy are described and analysed in Section 4. A
discussion of the present results and our conclusions are given in the
final section.

\section{Observational data}

The SDSS is based on observations obtained with a 2.5m telescope with
a 3$^\circ$ field of view at the Apache Point Observatory in New
Mexico, USA. Standard operations of the SDSS were performed from 2000
to 2005. Eventually, it will cover $\pi$ steradians centred on the
northern Galactic cap and three stripes near the southern Galactic cap
(Stoughton, Lupton \& Bernardi 2002). The operating mode of
observations is by means of drift scanning. One continuous scan is
designated by a ``run'', and two such runs generate a ``stripe''; they
have an overlap region of 1 arcmin on the edges of the two scans (Chen
et al. 2001). Catalogues of objects classified as ``stars'' from DR4
were downloaded from the sky server for sky areas selected to meet our
aims. The retrieved data is free of overlap and has been deblended;
saturated stars and poor-quality objects near bright stars and the
edges of the frames were all filtered out (Newberg \& Yanny 2005b).

If the Galaxy is axisymmetric, as usually assumed in previous studies,
the projected stellar number density at the same Galactic latitude
will increase towards the direction of $\ell=0^\circ$, and decrease
when approaching $\ell=180^\circ$, and any sky area pair
mirror-symmetrically selected on either side of the $\ell=0^\circ$
meridian plane should retrieve the same star number counts, within the
uncertainties. If the axisymmetric model is correct, such a picture
should be reproduced by star counts across the whole sky area, and can
in principle also be detected in symmetrically selected subsets of sky
areas. Limited by the current SDSS sky coverage, whole-sky star counts
are not possible, but a uniform data set covering all longitudes at a
given latitude should suffice, as this would be a perfect probe to
check the axisymmetric assumption. Such a data set is now available
from SDSS DR4, which will be addressed in the following. Fig. 1 is a
Lambert projection of the northern Galactic hemisphere, showing the
selected sky areas used in this work (marked by diamonds). Each of the
selected sky areas has a size of about $2^\circ \times 2^\circ$; the
Galactic coordinates of the centres of the selected fields are given
in Table 1. The sky areas selected along the circle at $b=60^\circ$
form the first group of our working data set. To check the structure
at latitudes different from $b=60^\circ$, sky area pairs at different
latitudes (from $b=55^\circ-85^\circ$, equally spaced by $5^\circ$)
along our longitudinal grid (in steps of $30^\circ$ from 0 to $\sim$
360$^\circ$) are also selected. In order to minimise the effects of
extinction, sky areas below $b=55^\circ$ were excluded. From the
present SDSS public data release, sky areas of 11 longitudinal strips
are available, and selected field from these form the remaining 11
data groups in our sample. From the second to the twelfth group, the
sky areas within each group have the same longitude, so that the
structural properties reflected by the stellar number densities along
a given longitude can be derived. In addition, through comparison
between sky areas having mirror symmetry with respect to the
$\ell=0^\circ$ meridian deviations from the axisymmetric Galaxy model
can also be checked. Owing to the SDSS observing strategy, stars
brighter than $15^m$ will be saturated, and star counts will be not
complete for magnitudes fainter than $22^m.2$ in the $g$ and $r$
bands, so that our work will be limited in the range of magnitudes
from $15^m$ to $22^m$ in both the $g$ and $r$ bands.

\section{The theoretical model}

To fully reveal the structure of the Milky Way, accurate measurements
of the distances to individual objects are key ingredients. Such
information is still lacking at the present time, however. As a
consequence, the structure of the Galaxy cannot be resolved
satisfactorily. Very large-scale photometric surveys greatly improve
our knowledge of Galactic structure. However, the problem remains even
if we use the data generated by the SDSS project, because of our
inability to determine stellar distances accurately. The major science
drivers of the SDSS are focused on extragalactic objects. Therefore, a
wide-filter photometric system was designed to detect the faintest
possible extragalactic objects. The narrowest filter, $u$, is
600{\AA}. Such a photometric system has an obvious disadvantage when
one wants to measure absolute stellar magnitudes, compared with the
classical $ubvy\beta$ system. The latter system by itself is already
characterised by a scatter of 0.3 mag for A and F-type stars (Crawford
1975, 1979). The difficulty to distinguish between giant and
main-sequence stars will affect counts of G and K-type stars even more
seriously in the SDSS system if one applies the colour-magnitude
relation for main-sequence star to all stars. Furthermore, the large
range of metallicity that is intrinsically present in the sample will
inherently lead to large scatter in the colour-magnitude relation. To
obtain distances to stars and to determine the structure of the Galaxy
directly from SDSS data is thus somewhat difficult. A preliminary
approach to this problem using the photometric parallax method and its
pitfalls are carefully analysed by Juri\'c et al. (2005).

To derive the structure of the Galaxy from the integrated stellar
light distribution along the line of sight, a theoretical model is
needed. Our Galactic disk models follows is based on that of Bahcall
\& Soneira (1980), while we have also included a thick-disk component.
Because of large deviations from an axisymmetric structure (described
in Section 4), we have modified the standard axisymmetric halo; these
deviations will be accounted for by a triaxial halo instead.

The basic star count equation we use is:
\begin{equation}
A(M_1,M_2, \ell, b) = \int_{M_1}^{M_2} {\rm d}M^{'} \int_0^\infty
R^2\,{\rm d}R\,\rho(\mathbf{r})\, \phi(M)\, {\rm d}\Omega,
\end{equation}
where $A(M_1,M_2, \ell, b)$ is the projected surface density in the
absolute magnitude bin from $M_1$ to $M_2$ in the direction of
Galactic longitude $\ell$ and Galactic latitude $b$; $R$ is the
heliocentric distance of a given star; $\rho(\mathbf{r})$ is the
density profile of each stellar population as a function of
$\mathbf{r}$, the distance to the Galactic Centre; $\phi(M)$ is the
luminosity function of each contributing stellar population, which is
a function of absolute magnitude and d$\Omega$ is the solid angle
element covered by the observations.

For the thin and thick disk components, the density profile is assumed
to be of the following exponential form,
\begin{equation}
\rho(\mathbf{r})=\exp[-|z|/H-x/h],
\label{eq_diskdensity}
\end{equation}
where $z$ is the height above the Galactic plane, $x$ the projected
distance to the Galactic Centre in the Galactic plane, and $H$ and $h$
are the scale height and scale length in $z$ and $x$, respectively. As
this study focuses on the structure of the halo, we used fixed models
for both the thin and the thick disks, so that complications due to
extra free parameters for these disks can be excluded. The parameters
of the disks are taken from Chen et al. (2001). For the thin disk,
$H=330$ pc and $h=2.25$ kpc; for the thick disk, $H=650$ pc and
$h=3.5$ kpc. The galactocentric distance of the Sun is $r_0=8.5$ kpc,
and the vertical distance of the Sun from the Galactic disk plane is
$z_0=27$ pc.

For the halo, a triaxial model is adopted to fit the observational
data.  A set of suitable coordinate frames is built following Newberg
\& Yanny (2005b), of which the schematic diagram is shown in
Fig. 2. Adopting the position of the Sun as the origin of the
coordinate frame, the direction from the Sun to the Galactic Centre is
taken as the $x$-axis. The y-axis is the direction perpendicular to
the $x$-axis in an anti-clockwise sense within the Galactic plane,
while the $z$-axis coincides with the direction towards the northern
Galactic pole. The $x'$ and $y'$ directions are parallel to $x$ and
$y$, but their origin is at the Galactic Centre instead; these axes
also define a right-handed coordinate system. Furthermore, $R$ is the
distance from a star to the Sun, $r$ the distance from the star to the
Galactic Centre and $t$ is the projection of $r$ on the Galactic
plane.  Therefore, $z=R \times \sin(b)+z_0$, $z_0=27$ pc,
$t=\sqrt{r_0^2+R^2-2 \times R \times r_0 \times \cos(b) \times
\cos(l)}$. The Galactic Centre coordinate frame is rotated in such a
way that its axes are settled right on the three axes of the triaxial
ellipsoid. The transformation between the cooordinate frames can be
separated into three rotational steps: first, fix arbitrarily one of
the three axes and rotate the other two. The rotation angles of the
three steps, denoted by $\theta$, $\xi$ and $\phi$, are defined when
fixing the $z$, $x$ and $y$ axes, respectively. In Fig. 2, for
example, where we rotate $x'$ and $y'$ around the $z'$ axis, the
relevant transformations are, $x_1=x' \times \cos(\theta)+y' \times
\sin(\theta)$, $y_1=-x' \times \sin(\theta)+y'\times \cos(\theta)$,
and $z_1=z'$. The remaining two steps involve doing the same for the
other two axes. The new axes obtained after the third rotation are
$(x_3,y_3,z_3)$, and the three axes of the triaxial ellipsoid are
$a,b$ and $c$, where $a$ is the longest and $c$ the shortest axis. The
axis ratios of the triaxial ellipsoid are $p=b/a$, and $q=c/a$, while
$r=\sqrt{x_3^2+(y_3/p)^2+(z_3/q)^2}$. The axisymmetric halo is
described by $p=1$.

A power-law (Reid 1993) is adopted for the halo density distribution,
\begin{equation}
\rho(\mathbf{r})=\frac{a_0^n+r_0^n}{a_0^n+r^n},
\end{equation}
where $a_0=1000$ is a normalisation constant. We adopt the $V$-band
luminosity function of Robin \& Cr\'e`\'e (1986) in our model, which
provides sufficiently accurate two-dimensional distributions in both
luminosity and spectral type down to $V=12$ mag. A two-dimensional
$V$-band luminosity function can easily be transformed to that in the
$g$ band. Robin \& Cr\'ez\'e (1986) compiled a number of luminosity
functions available at the time, including those of Wielen (1974,
1983) and Houk \& Cowley (1975, 1978, 1982). The work of Wielen is
widely used in star count studies to obtain number density-magnitude
relations for $2 \le M_V \le 12$ mag, while Houk \& Cowley's
luminosity function is often used for the bright stars ($-2 \le M_V
\le 2$ mag). Deutschman, Davis \& Schild (1976) and Hayes (1978)
provided calibrations of stellar luminosity and spectral type. To deal
with the data used for the present study, a further transformation
relation (Chen et al. 2001) from $V$ to $g$, $g=V+0.53(B-V)-0.075$,
was also adopted.

By using the density profile and the luminosity function discussed and
presented above, the distribution in absolute $g$ magnitude of the
stellar projected surface number density can be obtained from Eq. (1).

A colour-magnitude relation (CMR) is needed to obtain the number
density distribution in the $(g-r)$ colour. It is usual practice
to use the CMR defined by colour-magnitude diagrams (CMDs) of star
clusters.  However, there are usually too few stars in cluster
CMDs with absolute magnitudes fainter than $10^m$. Therefore, at
fainter magnitudes we do not have a precise CMR. As an alternative
approach, we adopt the theoretical isochrones of Girardi, Grebel
\& Odenkirchen (2004) to derive the CMR required here. The mean
ages of thin disk, thick disk and halo are assumed to be 4.5, 11
and 13 Gyr, and their metallicities are assumed to be,
respectively, 0.019, 0.004 and 0.0012. The theoretical CMR is
preferred because it includes faint stars that are otherwise not
available from observations. However, we have to bear in mind that
it cannot mimic the scatter on the relation in real situations.

Given a theoretical model for stellar distributions of different
Galactic components as described above, it is straightforward to
calculate the projected surface number density in absolute magnitude
bins in any given direction. From observations, the information that
can be directly extracted is the projected surface number density in
apparent magnitude bins. The goal of the present study is to use this
information to constrain the structure of the Galaxy parameterised in
the manner explained above.

A Monte Carlo method was adopted to simulate the observational
statistics.  Random points were generated based on several quantities,
including the luminosity function, distances weighted by the input
density profile, and a Gaussian-type observational error, so that the
observed star counts can be reproduced. For these simulations, we
adopted the three-dimensional extinction model of the Milky Way
derived from COBE observations (Drimmel, Cabrera-Lavers \&
L\'opez-Corredoira 2003). Instead of correcting the observational data
for the very complicated interstellar reddening, we applied reddening
corrections to the stellar distributions predicted by the model. The
extinction, $A_V$, is in general small throughout the areas selected
in this study, with a maximum of about 0.104 mag in the direction
(0$^\circ$, 55$^\circ$) and a minimum of 0.0385 mag in the
anti-Galactic Centre direction ($b=80^\circ$), at a distance of about
600 pc along the line of sight. $A_V$ is obtained through linear
interpolation for distances less than 600 pc. For greater distances,
$A_V$ is taken to be constant. The $g$-to-$V$-band extinction ratios
were taken from Girardi et al. (2004), and thus we can also obtain the
apparent $g$ and $r$-band magnitudes at each theoretical point.

We can now caluclate the theoretical projected surface number density,
theoretical star counts as a function of apparent $g$ and $r$
magnitude, and theoretical star counts as a function of $(g-r)$ colour
down to the SDSS limiting magnitudes, and hence we are now ready to
compare these with the observational data.

\section{Results and analysis}

\subsection{Analysis of the first group of observational data at $b$=$60^{\circ}$}

The discrete small sky areas of group 1, covering a complete circle at
the Galactic latitude of $b$=$60^{\circ}$, form a unique probe of the
structure of the Galaxy in the current SDSS data release. Direct
measurements of the stellar number density distribution within this
data set can tell us whether or not it is axisymmetryical.

We tested this idea by fitting the surface density distribution
along the circle with an axisymmetric halo. Fig. 4 shows the
projected surface number density distributions of the theoretical
model and the observational distribution for stars of $g, r$ $\in$
$[15,22]$ mag, where the diamonds represent the observational
projected stellar surface number density. To account for
observational fluctuations, each sky field was divided into four
subfields of $1^{\circ} \times 1^{\circ}$ (except for the area
[$200^{\circ}$,$60^{\circ}$], which is too small for this
procedure), so that the statistical fluctuations can be measured.
The resulting fluctuations over the average of the four subfields
are shown in the figure as error bars on the data points.  The
dashed line is the theoretical prediction for an axisymmetric
model. The dashed theoretical line was normalised by taking the
average values of the surface density at
($0^{\circ}$,$60^{\circ}$) and ($180^{\circ}$,$60^{\circ}$) from
the observational data. It is worthwhile to point out that the two
parts of the dashed theoretical line, i.e., that between $\ell =
0^{\circ}$ and $180^{\circ}$, and that between $180^{\circ}$ and
$360^{\circ}$, are not perfectly symmetrical because of
extinction. In addition, due to the non-uniform extinction, the
surface number density curves shown in Fig. 4 are not smooth. It
is evident that the axisymmetric model does not fit the
observations very well. Comparing the observational data with the
axisymmetric model prediction, it is obvious that the
observational data between $\ell = 0^{\circ}$ and $180^{\circ}$ on
the one hand, and between $\ell = 180^{\circ}$ and $360^{\circ}$
on the other are highly asymmetrical; the projected surface
density in the fields at $\ell$ $>180^{\circ}$ is systematically
higher than that in the fields mirror-symmetrically located on the
other side of the meridian. The axisymmetric theoretical value is
a little higher than derived from the observations for longitudes
from $0^{\circ}$ to $180^{\circ}$, and much lower than implied by
the observational data for $\ell = 180^{\circ}$ to $360^{\circ}$.
The deviations from the axisymmetric model are rather smooth; they
seem to represent a systematic shift from the theoretical model.

To find out what kind of stars are responsible for these asymmetric
statistics, we performed a number of tests. In Table 2 we list the
star counts of stars with $g,r$ $\in$ $[15,22]$ mag in the paired sky
fields selected symmetrically with respect to the $0^\circ$ meridian
plane, as well as the relative fluctuations (defined as [surface
density of $\ell_2$ - surface density of $\ell_1$]/[surface density of
$\ell_1$], where $\ell_1$ and $\ell_2$ are the Galactic longitudes of
the paired fields). In Table 3, faint stars with $g,r$ $\in$ $[19,21]$
mag are considered in the same way. They show the same trend as the
previous data set, resulting in an uneven distribution with an even
higher asymmetric ratio.

In Fig. 3, star counts in four colours, $(u-g)$, $(g-r)$, $(r-i)$
and $(i-z)$, in the fields towards the directions of
($90^{\circ}$, $60^{\circ}$) and ($270^{\circ}$, $60^{\circ}$) are
compared. The magnitude ranges are, respectively, $15 \le u \le
22$, $15 \le g \le 22$, $15 \le r \le 22$, $15 \le i \le 21$, and
$15 \le z \le 20$. In each of the colour distribution profiles
there are two peaks, as shown in Fig. 3 (the blue tail at $(u-g)
\sim 0.1$ is mostly due to white dwarf stars). The left-hand peak
is dominated by halo stars, while the one on the right is
dominated by thin disk stars, as inferred from the colours. This
is because most of halo stars are far more distant than the bulk
of the disk stars, and therefore only the luminous stars in the
halo (predominantly main-sequence stars) can be detected. For
stars on the main sequence, the intrinsically brighter ones have
bluer colors. In addition, the halo stars are generally more metal
poor, and hence tend to populate the blue peak. Based on a similar
argument related to the stellar metallicity, thin disk stars would
form the red peak. Therefore, it is clear that the excess of the
stellar number density in the field at ($270^{\circ}$,
$60^{\circ}$) with respect to that at ($90^{\circ}$, $60^{\circ}$)
is due to the halo component. The amplitude of the excess depends
on colour, as shown in Fig. 3; it is larger in $(u-g), (g-r)$ and
$(r-i)$ than in $(i-z)$. It may be caused by the different
limiting magnitudes, or it might be an indication of the specific
properties of the stars that causes the excess.

\subsection{Fitting the projected surface number densities with a triaxial halo model}

A triaxial halo model, as described in Section 3, is adopted to
describe the asymmetry found in our data set. Given the same limiting
magnitudes as those of the observations, the theoretical surface
densities are calculated using a model of which the structural
parameters are constrained by a comparison with the observations.

The three-component model (including the thin and thick disks, and the
halo) is used to calculate projected surface number densities in the
sky areas corresponding to those of the observations. As the main
objective of the current work is to derive the structure of the halo,
for the parameters of the thin and thick disk we adopt the results of
Chen et al. (2001). The parameters of the halo are adjusted in such a
way that the observational data are best fitted. The integration of
the observational surface density over all of the selected fields is
used to constrain the surface density of the model.

There are a total of six parameters in our halo model, namely the
power-law index $n$, the two axis ratios $p$ and $q$, and the three
coordinate rotation angles $\xi$, $\phi$, $\theta$ (based on which the
$x, y, z$ axes of the triaxial ellipsoidal model can be translated to
that of the Galaxy). The possible ranges of the parameters are
estimated using initial tests over a preliminary grid, which is given
in Table 4.

Chi-squared minimisation tests comparing the theoretical and
observational data sets were carried out using the algorithm of Press
et al. (1992). For a non-Gaussian distribution of discrete data
points, Pearson's $\chi^2$ can be expressed as,
\begin{equation}
\chi^2=\sum_{i=1}^N \frac{(R_i-S_i)^2}{(R_i+S_i)(N-m)},
\label{eq4}
\end{equation}
where $R_i$ is the theoretical frequency of data points in the $i^{\rm
th}$ bin, and $S_i$ is the equivalent frequency of the
observations. $N$ is the number of bins, $m$ the number of model
parameters, and therefore $(N-m)$ is the number of degrees of
freedom. Assuming that the errors in the theoretical and observational
frequencies are distributed in a Poissonian fashion, we obtain
$\sigma^2=(\sqrt{R_i^2}+\sqrt{S_i^2})^2$. By changing the model
parameters, the parameter combination that generates the minimum
averaged $\chi^2$ of the 12 groups can be obtained.

Models for all possible combinations of the six parameters were
calculated, and the corresponding $\chi^2$s were evaluated to find the
parameter combination resulting in the minimum $\chi^2$ value. It is
obvious that the models with $(p,q)$ in the range of
$(0.5,0.5),(0.6,0.5),(0.7,0.6)$, and $(0.8,0.6)$ lead to better values
for $\chi^2$, regardless of any fine-tuning of the other four
parameters.  The models provide only weak constraints for the
power-law index $n$. When $n$ is fine-tuned, the axial ratio should be
adjusted accordingly, in order to obtain the optimum $\chi^2$. The
overall trend is that for larger $n$, larger $p$ or $q$ ratios are
needed.

Figs 4 and 5 show the observational surface densities in the selected
regions and those of our theoretical models calculated by using a
triaxial halo with the following parameter combination, $n=2.2, p=0.5,
q=0.5 ,\theta=60^\circ, \xi=-10^\circ$, and $\phi=-10^\circ$, and the
corresponding $\chi^2$ is 2.02. (This is the parameter set that leads
to the minimum $\overline{\chi^2_{g-r}}$, see Section 4.3 for more
details).  Fig. 4 shows the results for the group-1 fields, spanning a
circle at $b=60^{\circ}$. A comparison between the solid and dashed
lines in Fig. 4 shows that the triaxial halo model (the solid line)
fits the observations much better than an axisymmetric halo (the
dashed line), with the triaxial halo clearly showing the degree of
asymmetry as revealed by the observations. The only sizable deviations
of the triaxial model from the observations are found near
$\ell=260^{\circ}$ (where the theoretical prediction for the surface
density is slightly higher), and near $\ell=330^{\circ}$ (where the
theoretical value is somewhat lower). The six panels in Fig. 5,
labelled A-F, show the surface densities of both the observations
(diamonds and triangles) and the models (dotted and dashed lines) for
groups 2 to 12. Panel A shows the surface densities in the fields at
$\ell$=$0^{\circ}$ and $\ell$=$180^{\circ}$ as a function of Galactic
latitude. Panels B through E show the surface densities in the
sky-field pairs mirror-symmetrically selected on either side of the 0
degree Galactic meridian plane. Panel F depicts that in the
$\ell$=$150^{\circ}$ field, the companion field of which is not shown
because the data has not yet been released. From these figures, it is
clear that there are larger surface densities in the sky areas between
$\ell = 180^{\circ}$ and $360^{\circ}$ than in their corresponding
mirror-symmetric fields. These fields thus show the same trend as the
sky areas in the first group, at $b=60^\circ$. The theoretical model
fits the observations quite well in general, with several
exceptions. For the three groups at $\ell$=$0^{\circ}, 240^{\circ}$,
and $270^{\circ}$, the values are higher than those of the
observations around a latitude of $b$=$55^{\circ}$. For the group at
$\ell=300^{\circ}$ and $\ell$=$270^{\circ}$, the theoretical values
are lower than the observational value around $b=70^{\circ}$. For the
group at $\ell=330^{\circ}$ and $\ell$=$240^{\circ}$, the theoretical
values are lower than the observational values around $b=75^{\circ}$
and $b=55^{\circ}$, respectively.

In this paper, a power-law model is used to interpret the observations
for the purpose of clarity. Models implemented with more complex
physics, such as the Hernquist model (Newberg \& Yanny 2005b), can fit
the observations better. This could be one of the reasons that our
model does not fit the observations perfectly. Modifications to this
problem will be considered in a future study.

\subsection{Fitting star counts in the $(g-r)$ colour}

In the previous section, the projected surface number densities were
fitted using a triaxial halo model. Star counts combined with colour
information can, however, reveal more information about the stellar
populations in the data and constrain the model better. Therefore, we
performed model fits in the $(g-r)$ colour. The model described in
Section 3 is now used to calculate star counts in the $(g-r)$ colour.

By using Eq. (1), the projected stellar number density in each
absolute magnitude bin can be obtained. Monte Carlo simulations were
carried out for absolute magnitude and distance, so that artificial
stars could be generated for a given model, and hence theoretical
apparent magnitudes could be generated. The $(g-r)$ colour as well as
its statistical uncertainty at each point can then be obtained from
the colour-magnitude relation.

Comparisons between the theoretical and observational star counts
for all the fields were also done following Eq. (4). Subsequently,
the $\chi^2_{(g-r)}$ values of all fields in the 12 groups were
averaged to obtain the mean value $\overline{\chi^2_{(g-r)}}$.
Intensive and elaborate computational tests were done for all
possible cases. Each set of parameters represents a model which
can be evaluated by its $\chi^2$ and $\overline{\chi^2_{(g-r)}}$.
It was found that $\chi^2$ and $\overline{\chi^2_{(g-r)}}$ exhibit
similar trends. However, when $\overline{\chi^2_{(g-r)}}$ reaches
its minimum value, $\chi^2$ may not be at its minimum. Therefore,
we have to sacrifice minimizing the latter in order to keep
$\overline{\chi^2_{(g-r)}}$ small. Table 5 lists the best-fit
parameter combinations for power-law indices $n$ from 2 to 2.6.
Because of the six-dimensional parameter space, multiple solutions
were derived. Fitting actual star counts in $(g-r)$ is much more
difficult than fitting the projected surface number density. The
minimum $\chi^2$ value is about 1.735, but that of
$\overline{\chi^2_{(g-r)}}$ is a few times higher. In our model,
the power-law index $n$ still leaves much room for fine-tuning,
from 2.0 to 2.6 (beyond which $\chi^2$ and
$\overline{\chi^2_{(g-r)}}$ become too large). If constrained only
by the projected number density (see Section 4.2), no firm
constraint on the axial ratios $(p,q)$ can be reached. When
supplemented with star counts in $(g-r)$, each $n$ has only one
best-fit $(p,q)$ combination. For $n\leq 2.4$, $(p,q)$ is $(0.5,
0.5)$. This best-fit parameter set is different from that of the
traditional axisymmetric model, even if $p=q$. In this model, the
y axis is equal to the z axis, and therefore resembles the
rotationally symmetric model, but with an inclination and tilt
with respect to the Galactic plane. We would like to call all
these $p=q$ cases also triaxial models. In such cases, good fits
to the observational data can also be achieved. However, the
ellipticity seems too oblate. When $n=2.4$, $(p,q)$ becomes
$(0.6,0.5)$; when $n=2.5$ or 2.6, $(p,q)$ goes to $(0.7,0.6)$. The
larger $n$ becomes, the higher the axial ratio will need to be.
Directly following from the power-law distribution, the number
density decreases more rapidly for larger $n$ when going away from
the Galactic Centre. A higher axial ratio tends to enhance the
number density in the directions of the axes. Hence, for larger
$n$, a larger axial ratio is required to balance the effect of the
larger $n$ in order to fit the observational data. Such a
correlation is the same as that for the direct surface number
counts in Section 4.2.

Figs 6 and 7 show star counts in $(g-r)$ for the $b$=$60^{\circ}$
fields. The theoretical values were calculated with the same model
as in Figs 4 and 5. Our purpose is to compare the mirror-symmetric
pairs; four pairs of fields at approximately this latitude are
considered. Because the sky field at ($30^{\circ},60^{\circ}$) is
not included in the survey, we chose the field at
($30^{\circ},65^{\circ}$) to be paired with that at
($330^{\circ},65^{\circ}$). The thin-line histogram represents the
observational data, the solid black histogram describes the
theoretical results. In Fig. 6, panel A shows the results for the
sky field at ($0^{\circ},60^{\circ}$), and panel B for
($180^{\circ},60^{\circ}$). The panel pairs C-D, E-F, G-H and I-J
are for four pairs of mirror-symmetric sky fields. Panel K in Fig.
7 is for the field at ($150^{\circ},60^{\circ}$), a single field
without a companion on the other side of the Galactic meridian
plane.  Comparing the observational data in the paired sky fields,
the histograms for $\ell$ from $180^{\circ}$ to $360^{\circ}$
exhibit higher halo star peaks, and similar peaks for the thin
disk with respect to those on the other side of the meridian. The
theoretical model follows the same trend as the observations.

Apart from the problem of the structural parameters, the stars in the
Galaxy are simplified as three populations (thin and thick disks, and
halo), and it is further assumed that each population internally has
the same age and metallicity. All of these assumptions will surely
affect the distribution of the stars in the $(g-r)$ colour. A more
precise $g$-band luminosity function and better partitioning of the
populations are needed to interpret the continuously updating
large-scale survey data.

\section{Discussion and Conclusion}

Star counts based on the observational data of the Milky Way analysed
in the present paper show an asymmetry in the large-scale areas of the
sky surrounding the northern Galactic pole, and therefore they reflect
a certain non axisymmetric structure. These observations may be
explained in one of two ways. The present smooth Galactic halo itself
may have an asymmetric structure, which can be represented by a
triaxial ellipsoid, possibly tracing the dark halo gravitational
potential that might be triaxial (Jing \& Suto 2002). The triaxiality
proposed as such is not unique; it is considered common for dark
haloes in extragalactic systems. Mazzei \& Curir (2001) analysed the
effects of a triaxial dark halo on the bar-triggered star formation
and feedback processes after the formation of the disk. They indicated
that the dark halo has important effects on the evolution of baryonic
matter. The star-formation rate is not only linked to the total mass
of the dark halo, but also to the dynamical state of the dark halo. Is
it then normal to form an asymmetric stellar halo in such an
asymmetric gravitational field? In the history of formation and
evolution of the Galaxy, the halo experienced accretion and merger
processes. Wyse \& Gilmore (2005) claim that merger events and
accretion of satellite galaxies in the past history of the Galaxy do
play some role in the evolution of the Milky Way. The outer halo, in
particular, might be dominated by substructures that are likely the
remnants of interactions. If the halo were very sparse originally and
were gradually built up from accretion and merger events, it would be
expected that there have been accretion events happening throughout
the lifetime of the Milky Way. This would allow for any sizable
features from early times to be smoothed out via dynamical evolution,
leading to the formation of a smooth structure at the present
time. Presumably, the star streams should have a random distribution
due to the random infall angle and intensity, and therefore -- on
average -- the overall features would have a smooth distribution after
relaxation. Newberg and Yanny (2005b) selected F-type turn-off stellar
data in eight SDSS stripes from the DR3 to map the distribution of
stars in real space. They selected fields that most likely contain
only halo stars. Sky areas near any previously identified
overdensities caused by star streams were also avoided in their
studies, but their results can be altered by the Virgo overdensity
discovered later by Juri\'c et al. (2005). After they compared the
observational data with the results of five triaxial models, they
obtained the best-fit parameter set for an oblate ellipsoid with its
major axis pointed at 50--$70^{\circ}$ from the line of sight towards
the Galactic Centre from the Sun, and with a 4--6$^\circ$ inclination
with respect to the Galactic plane. The minor axis, z, has a length
equivalent to 65\% of the major axis. The intermediate axis, y, is
about 75\% of the major axis (Newberg \& Yanny 2005b).

An alternative cause of an asymmetric halo may be found in large-scale
star streams that are embedded in the smooth background and that have
not yet been melted into the halo completely. Also using DR4 data,
Juri\'c et al. (2005) analysed the structure of the Galaxy using the
photometric parallax method. They found that there are significant
overdensities in the range ($5 < z/{\rm kpc} < 15$), with the
overdensity peaking in the direction of Virgo, and the entire
overdensity region covers about one thousand square degrees on the
sky. They selected both a sky field in the overdensity region and a
control sky field mirror-symmetrically located on the other side of
the meridian with respect to the target field, as shown in figure 24
of Juri\'c et al. (2005). By comparing stars with $0.2 < (g-r) < 0.8$
mag and $18 < r < 21.5$ mag in both fields, they estimated the surface
brightness and luminosity of the overdensity in the target field as
$\sum_r=32.5$ mag arcsec$^{-2}$, and $L_r=0.09 \times 10^6$
L$_\odot$. Based on this result, we can calculate the additional
number of stars per square degree due to the overdensity. The
magnitude range used by Juri\'c et al. (2005) spanned from $r = 18$ to
21.5 mag. Adopting the distance to the overdensity as 10 kpc, as also
assumed by Juri\'c et al. (2005), the lower and upper limit of the
$r$-band absolute magnitude of their sample stars is therefore from
6.5 to 3 mag. From the transformation relation $r=V-0.49(B-V)+0.11$
(Chen et al. 2001), the limiting magnitudes in the $V$-band can be
obtained (taking the $V$ -- $(B-V)$ relation for main-sequence stars
from Cox 1999). The Mass -- $V$ magnitude relation from Reid (2002),

\begin{equation}
\log M=0.477-0.135M_V+1.228\times10^{-2}\times
M_V^2-6.734\times10^{-4}M_V^3 ,
\end{equation}
is used to calculate the mass of the most luminous (1.45
M$_\odot$) and the faintest stars (0.81 M$_\odot$) corresponding
to the respective observed magnitude thresholds. Integrating the
luminosity-mass ratio, weighed by $m\times$ IMF (Reid 2002), the
additional stellar density number per square degree can be
calculated,

\begin{equation}
N = L_{\rm tot}/(c \int (L/m)m \phi(m){\rm d}m ),
\label{eq_number}
\end{equation}
where $c$ is the normalisation constant of the IMF. The derived number
is 614 stars per square degree, which is indeed enough to match the
number density excess with respect to a symmetric structure.  However,
this is still not enough to fully resolve the problem. By subtracting
the CMD of a control field from that of the sky area characterised by
the density excess, the flux of the Virgo star stream was
derived. However, if a triaxiality does exist in the structure, there
would be no proper criteria to distinguish the contribution due to the
star stream from that of triaxiality.

As demonstrated in Fig.~4, there is a large region of stellar
underdensity with respect to an axisymmetric model, which is centred
at about $\ell=150^\circ$ in the direction of Ursa Major (UMa) with an
amplitude less than the Virgo overdensity. This was also shown in
figure 3 of Newberg \& Yanny (2005a), where they quote a minimum at
$\ell=155^\circ$ for F-type turn-off stars, very close to what we have
found in this paper. The size of the UMa underdensity region is
smaller than, but comparable to, that of the Virgo overdensity. The
UMa underdensity is also clearly visible from figure 24 of Juri\'c et
al. (2005; the blue colour-coded region near $\ell=150^\circ$), and
more prominently shown on the colour map in a recent news release on
the SDSS web page. Such an underdensity region can also be found in
the spatial distribution of M-giants from 2MASS observations (figure
27 of Juri\'c et al. 2005). The Virgo overdensity and the UMa
underdensity regions are well aligned. If the overdensity is to be
accounted for only by a star stream, what is then the cause for the
observed underdensity aligned with it? This infers that there is some
substructure within the halo that cannot be accounted for only by a
star stream such as that of the Virgo overdensity. As shown in Fig. 4,
the overdensity is observed in a very large region, from $\ell =
180^\circ$ to $360^\circ$, which is more than a quarter of sky at the
latitude of $b = 60^\circ$. It cannot be fully understood by a stream
covering only 1000 deg$^2$. The triaxial halo model may provide a
plausible explanation, because the distribution of the stellar surface
number density at $b=60^\circ$, including these over- and
under-densities, can be fitted simultaneously by a triaxial halo
model.

To disentangle the effects of a triaxial halo and the large-scale star
stream, kinematic and chemical information is needed.

The substructures detected in the SDSS data are summarised in Table 2
of Newberg, Yanny \& Rockosi (2002); two of their star streams pass
through the sky fields selected in this study, namely S341+57$-$22.5
and S297+63$-$20.0. Since S341+57$-$22.5 is below the magnitude limits
of the survey, it can hardly alter our results.  S297+63$-$20.0 may
have a minor influence on our results because of its limited size in
the given direction. However, it will not change the general
large-scale picture of the current work.

As the primary goal of the current work, a study of the large-scale
structure of the Galactic halo using star counts is presented.  The
star streams can not easily be corrected for from the SDSS
data. Therefore, we chose all-sky coverage to try to fit the average
structure. We have shown that the stellar projected surface density of
the halo stars at $\ell$ from $180^{\circ}$ to $360^{\circ}$ is
significantly higher than that at $\ell$ from $0^{\circ}$ to
$180^{\circ}$ at a latitude of 60$^\circ$ below the SDSS magnitude
limit. The observed stellar projected surface density distributions in
the sky fields selected for the current work can be well fitted by
using a triaxial halo model. The parameter combinations of the
best-fit triaxial models are, power-law index $n$ from 2.0 to 2.6,
$\theta$ from $55^{\circ}$ to $65^{\circ}$, $\phi$, $\xi$ less than
$5^{\circ}$, $(p, q)$ either $(0.5,0.5)$, $(0.6,0.5)$, or $(0.7,0.6)$.
These parameter sets agree with the one of Newberg \& Yanny (2005b),
except that their axial ratios are larger than ours. The results of
the current work can still be affected by lumps in the star streams
that have not yet been smoothly dispersed into the halo.

We are left with a number of questions, such as: what are the bulk of
the stars in the star streams? What is the proportion of stars in the
streams across the whole halo? Do star streams contribute
significantly to the formation of a triaxial halo, or do they simply
act as minor disturbances to the overall profile? How many of the star
streams have already dispersed into the halo and become a smooth
background, and how many of them can still be picked up from the
smooth halo background? There is no way to have any clear answer for
these questions from the present data. Work in this subject will leap
forward with the coming on-line of future projects aimed at
spectroscopic sky surveys such as SEGUE, LAMOST, SDSS II, GAIA, and
further photometric surveys of the southern sky.

\section*{Acknowledgments}
We would like to thank B. Chen, Y. C. Liang, S. D. Mao, Z. G.  Deng,
X. Y. Xia, C. H. Du, Y. B. Yang, and Y. G. Wang for valuable
suggestions and fruitful discussions. We are also very grateful for
email communications with H. J. Newberg and A. Robin. We would like
to thank our colleague, Dr. R. de Grijs, for proof reading the
paper. This work is supported by the National Natural Science
Foundation of China under grant No. 10333060, No. 10573022 and No.
10403006.

\begin{figure*}
\includegraphics[scale=0.8]{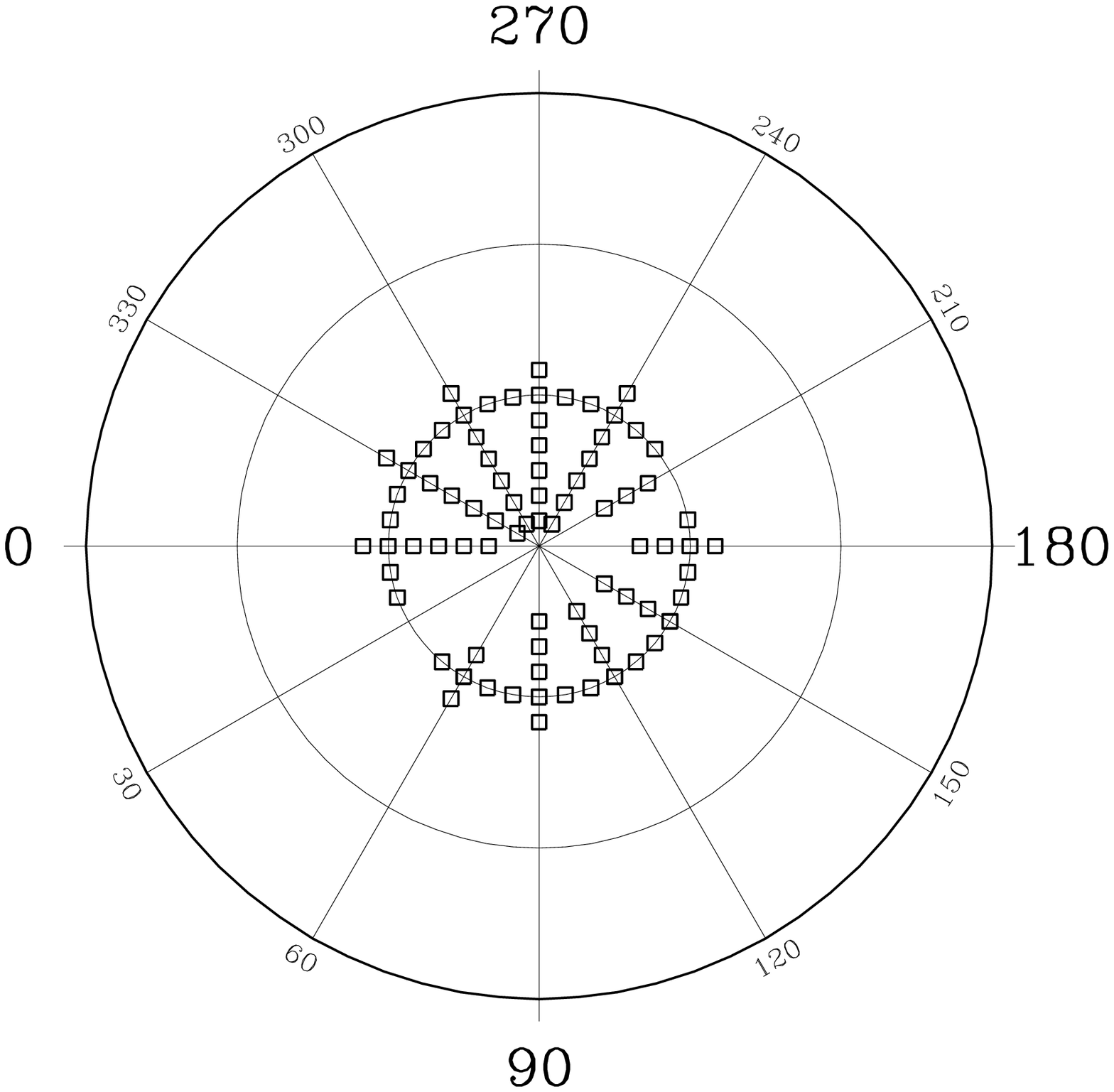}
\caption{A sketch of the selected sky areas of the current study in
Galactic coordinates.\label{skycover}}
\end{figure*}

\begin{figure*}
\includegraphics{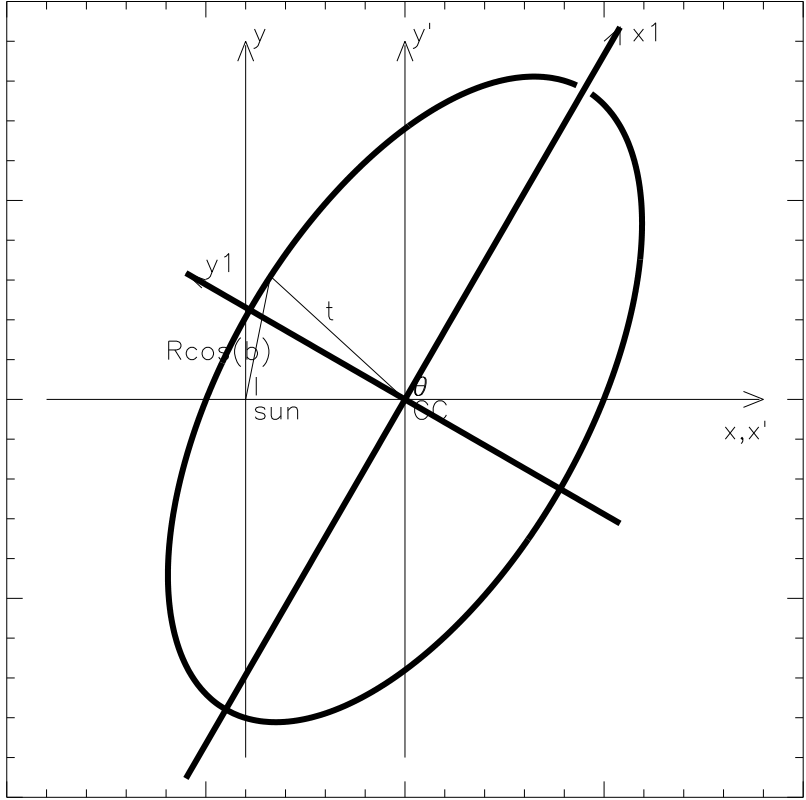} \caption{A sketch of the orientation of the coordinate
frames.\label{triaxial}}
\end{figure*}

\begin{figure*}
\includegraphics{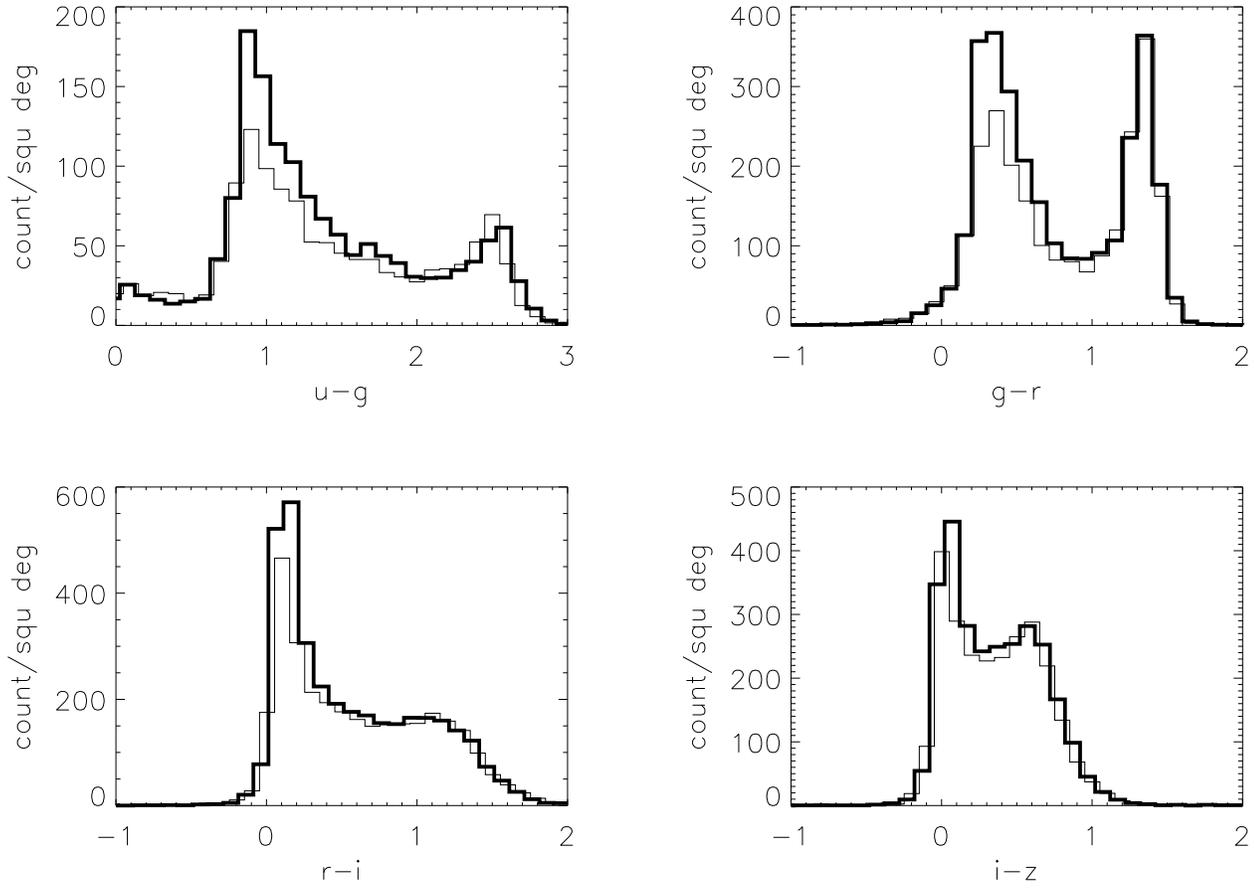}
  \caption{\label{colc90vs270}
Star counts in four colors, namely $(u-g), (g-r), (r-i)$ and $(i-z)$. The
thick solid histogram represents the star counts at ($270^{\circ}$, $60^{\circ}$),
and the thin histogram describes star counts at ($90^{\circ}$,
$60^{\circ}$); the two sky fields were chosen in mirror-symmetry with
respect to the of l=$0^{\circ}$ Galactic meridian plane.}
\end{figure*}

 \begin{figure*}
  \includegraphics[scale=0.8]{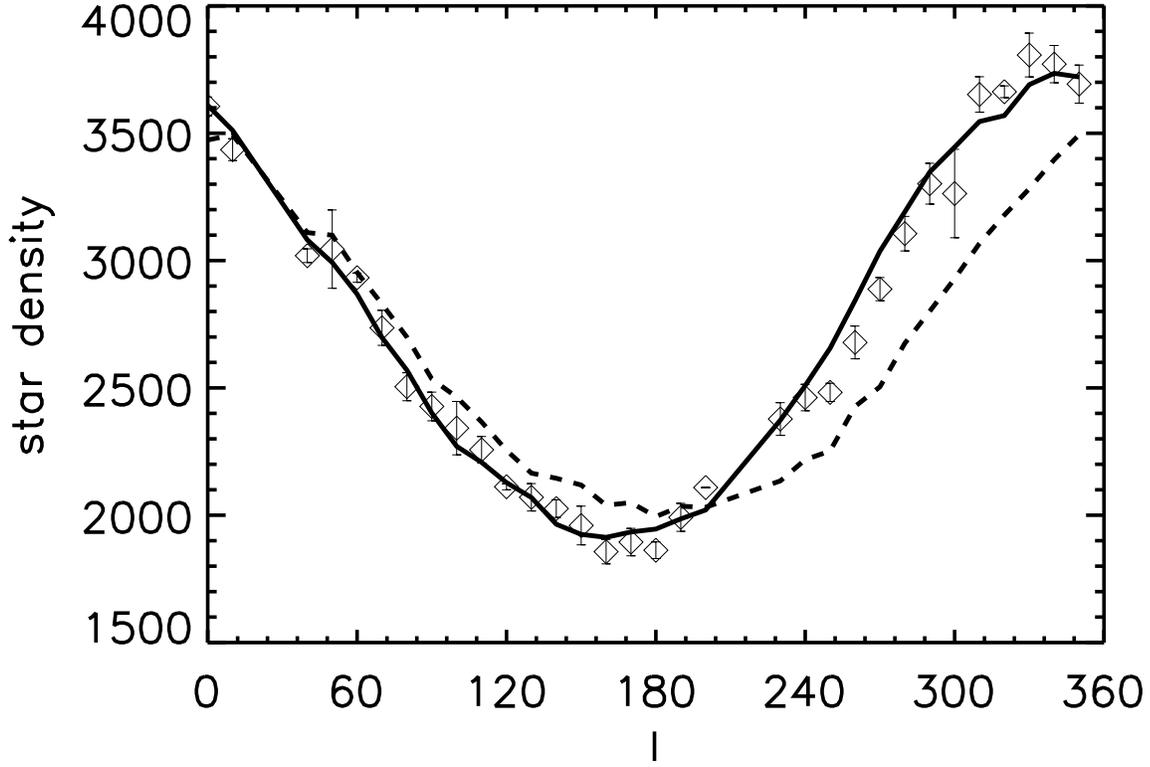}
  \vspace{20pt}
  \caption{\label{f:compare60}
Stellar surface number density distribution at a latitude of
60$^\circ$. The diamonds describe the surface number density of
the observational data with $g$ and $r$-band magnitudes from 15 to 22
mag; the error bars represent density fluctations. The dashed line
describes the theoretical result calculated based on an axisymmetric halo
model, and the solid line describes the theoretical result for triaxial
halo model.
  }
\end{figure*}

\begin{figure*}
  \includegraphics[scale=0.4]{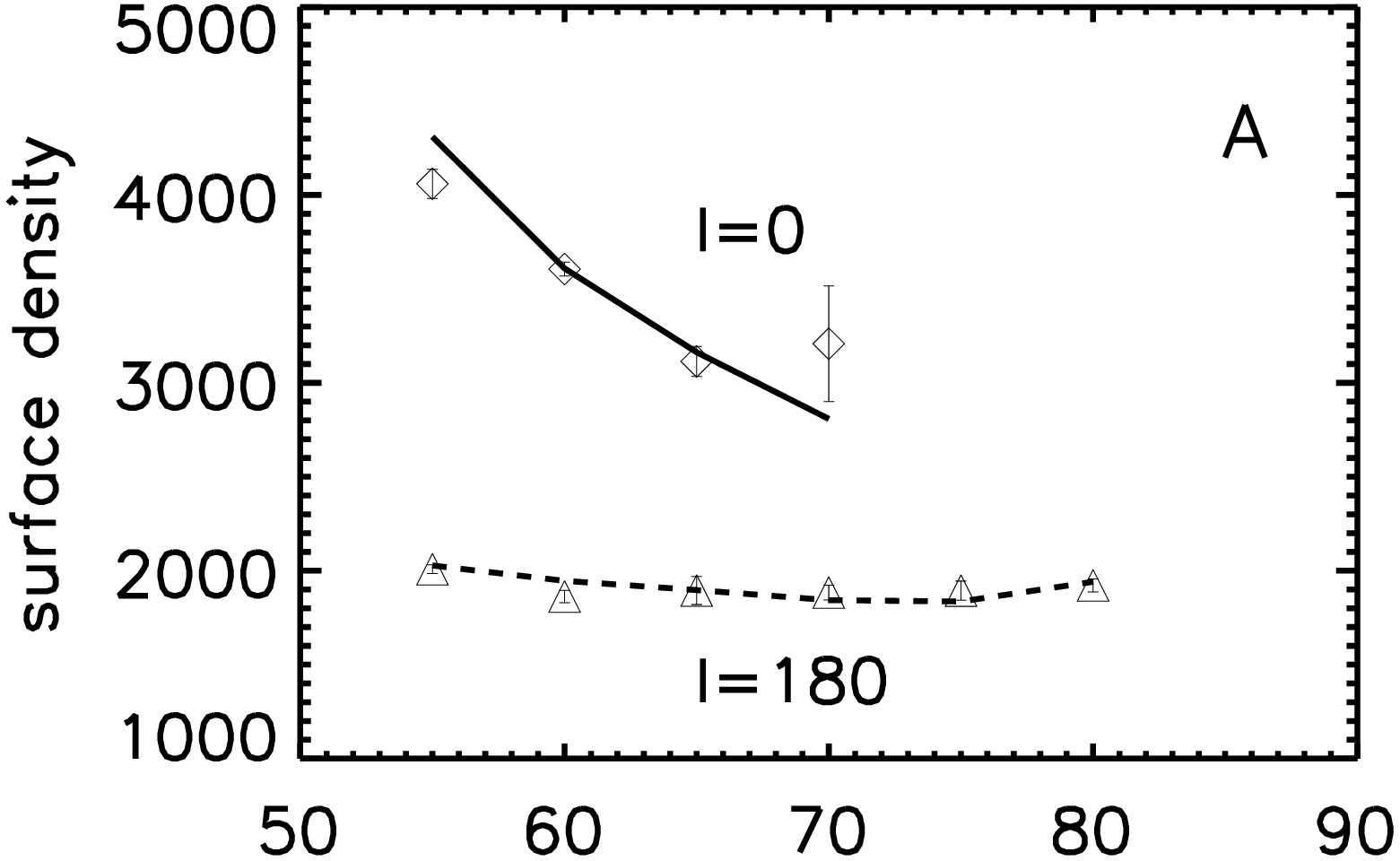}
  \hfill
  \includegraphics[scale=0.4]{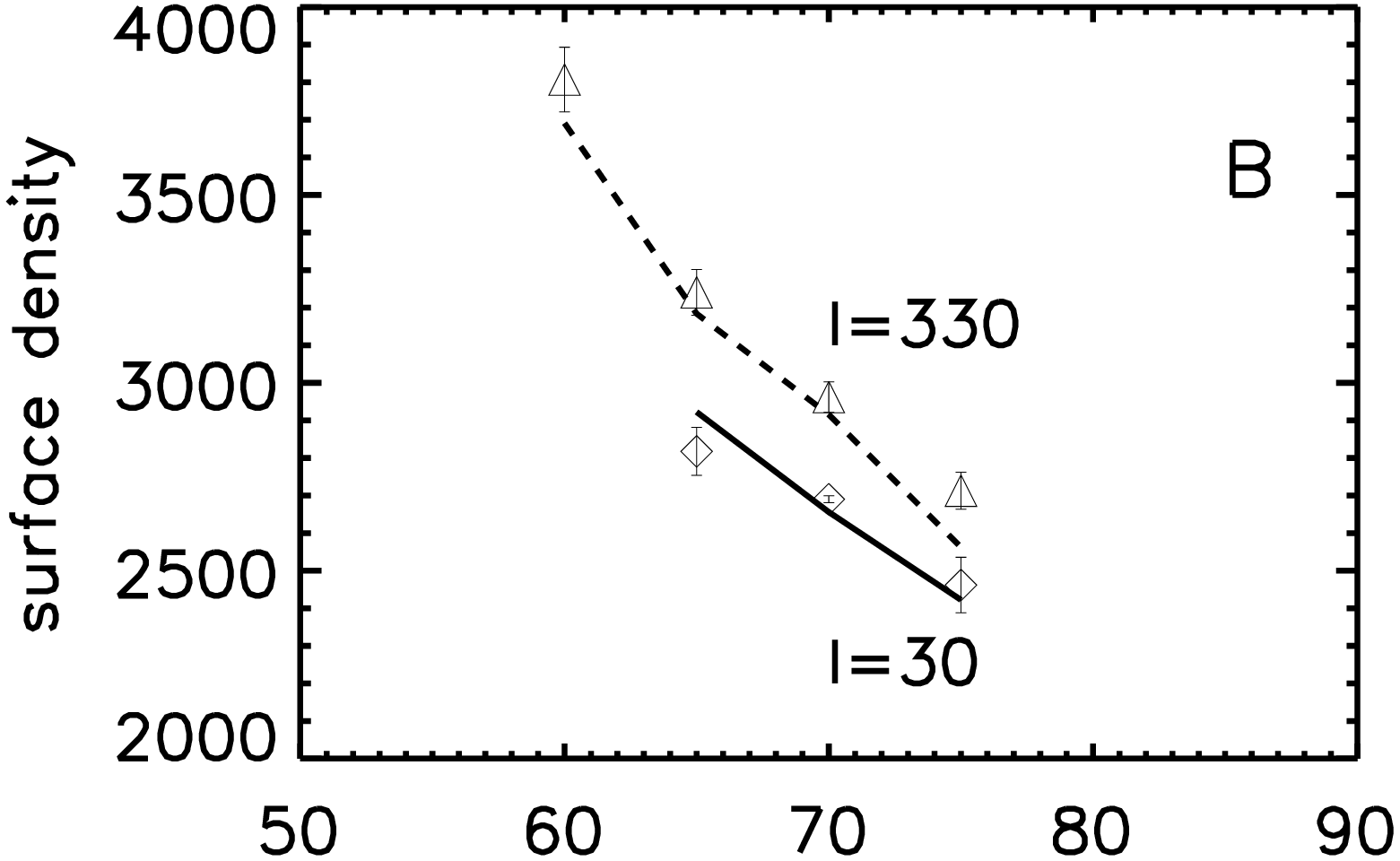}
  \vfill
  \includegraphics[scale=0.4]{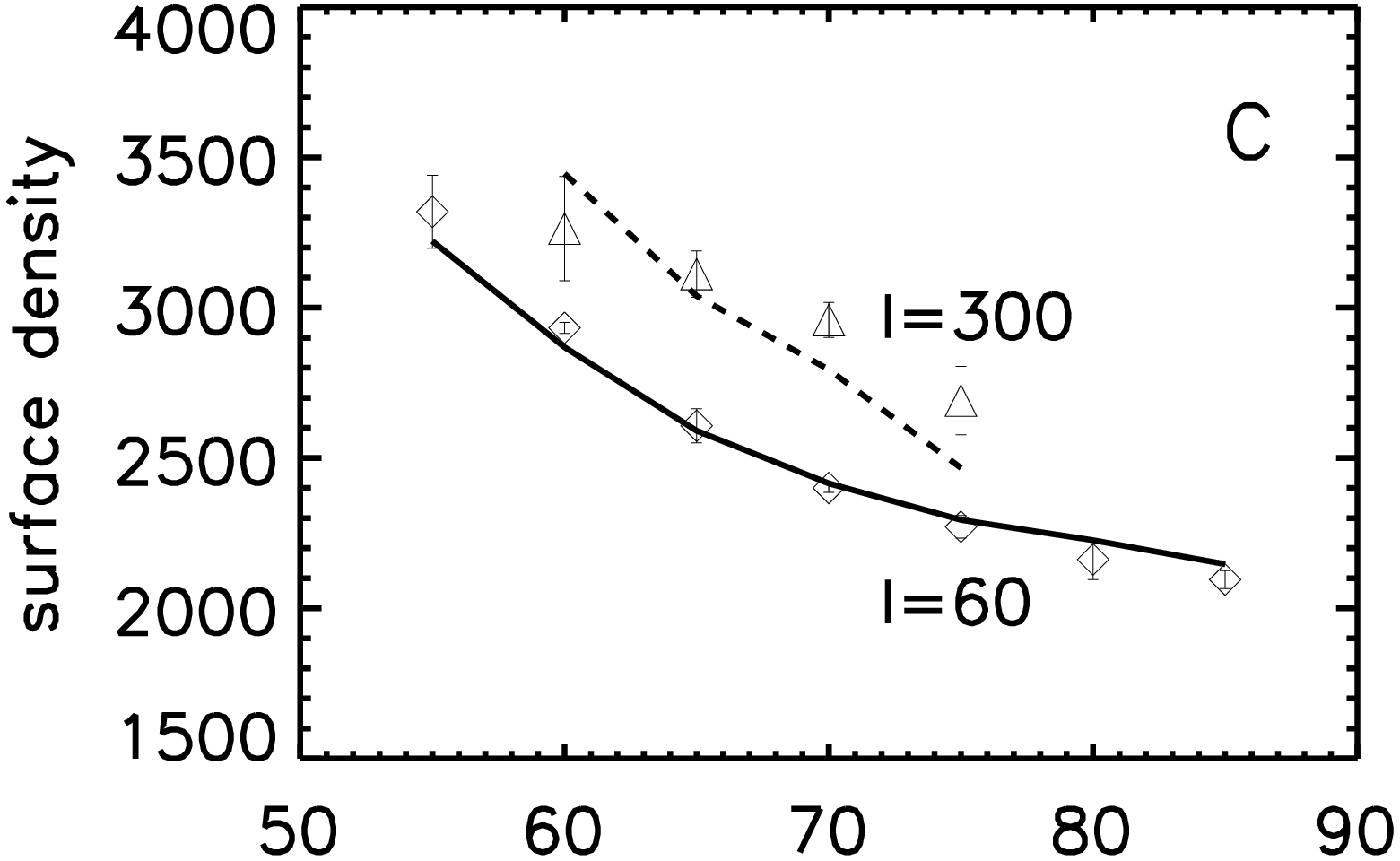}
  \hfill
  \includegraphics[scale=0.4]{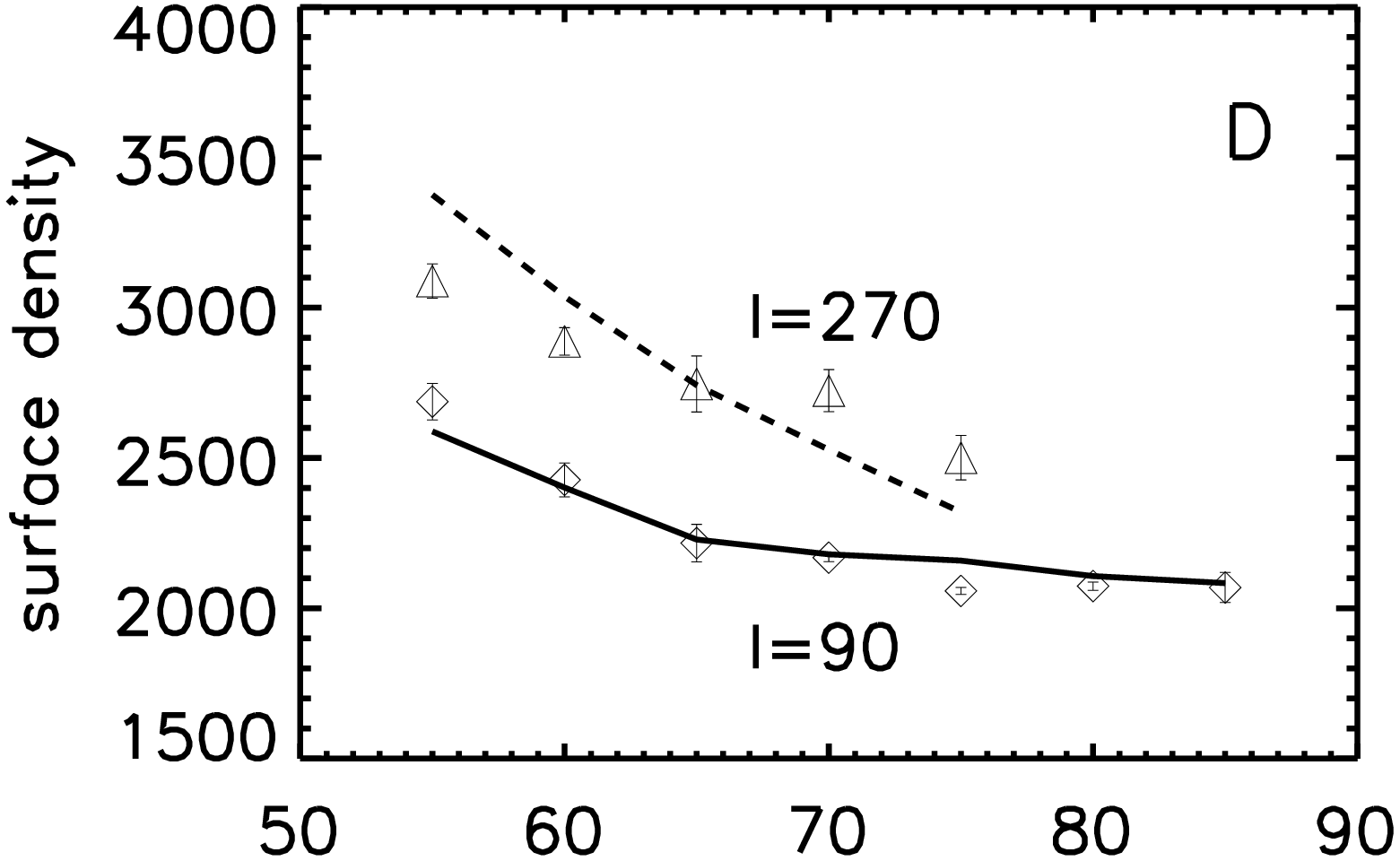}
  \vfill
  \includegraphics[scale=0.4]{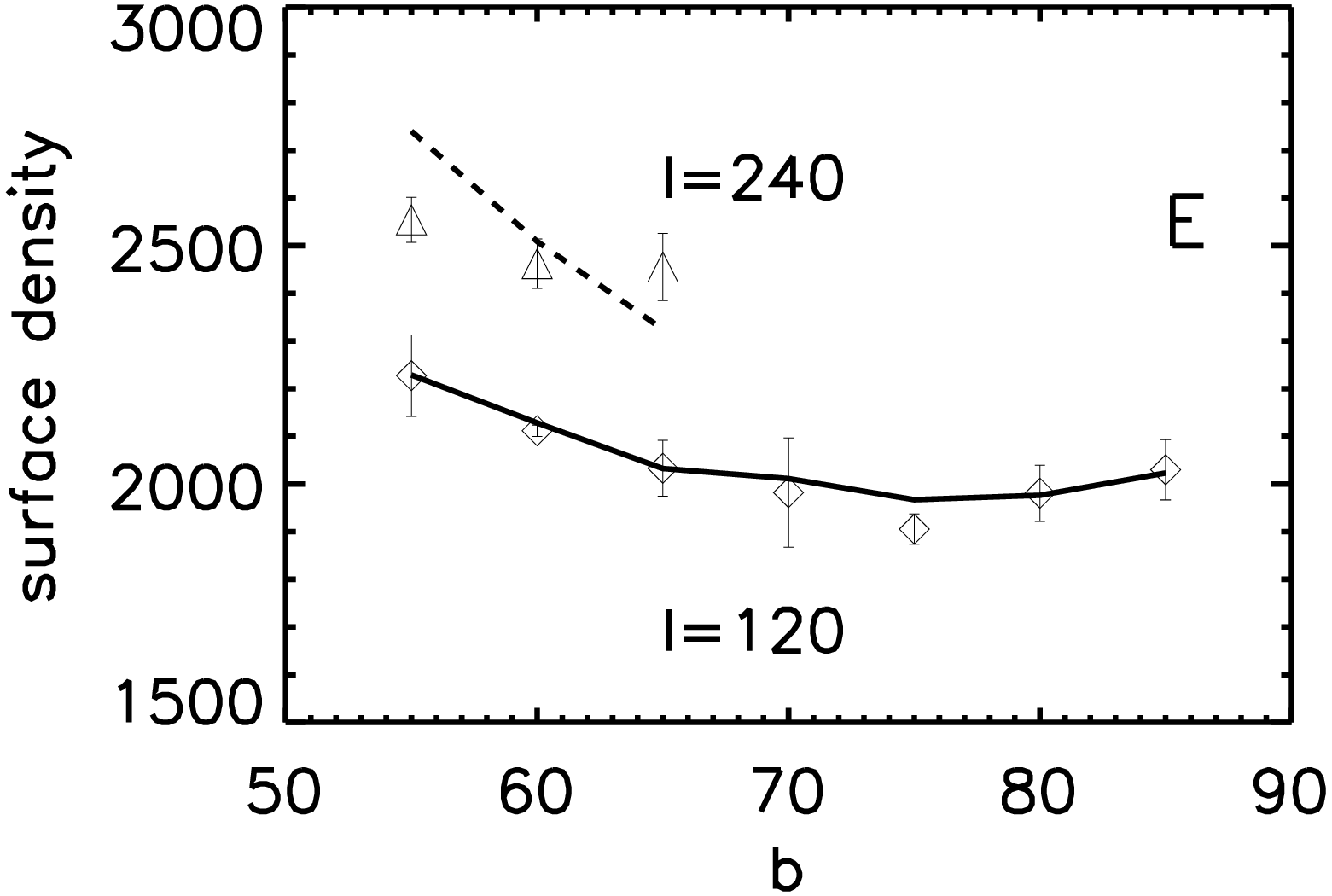}
  \hfill
  \includegraphics[scale=0.4]{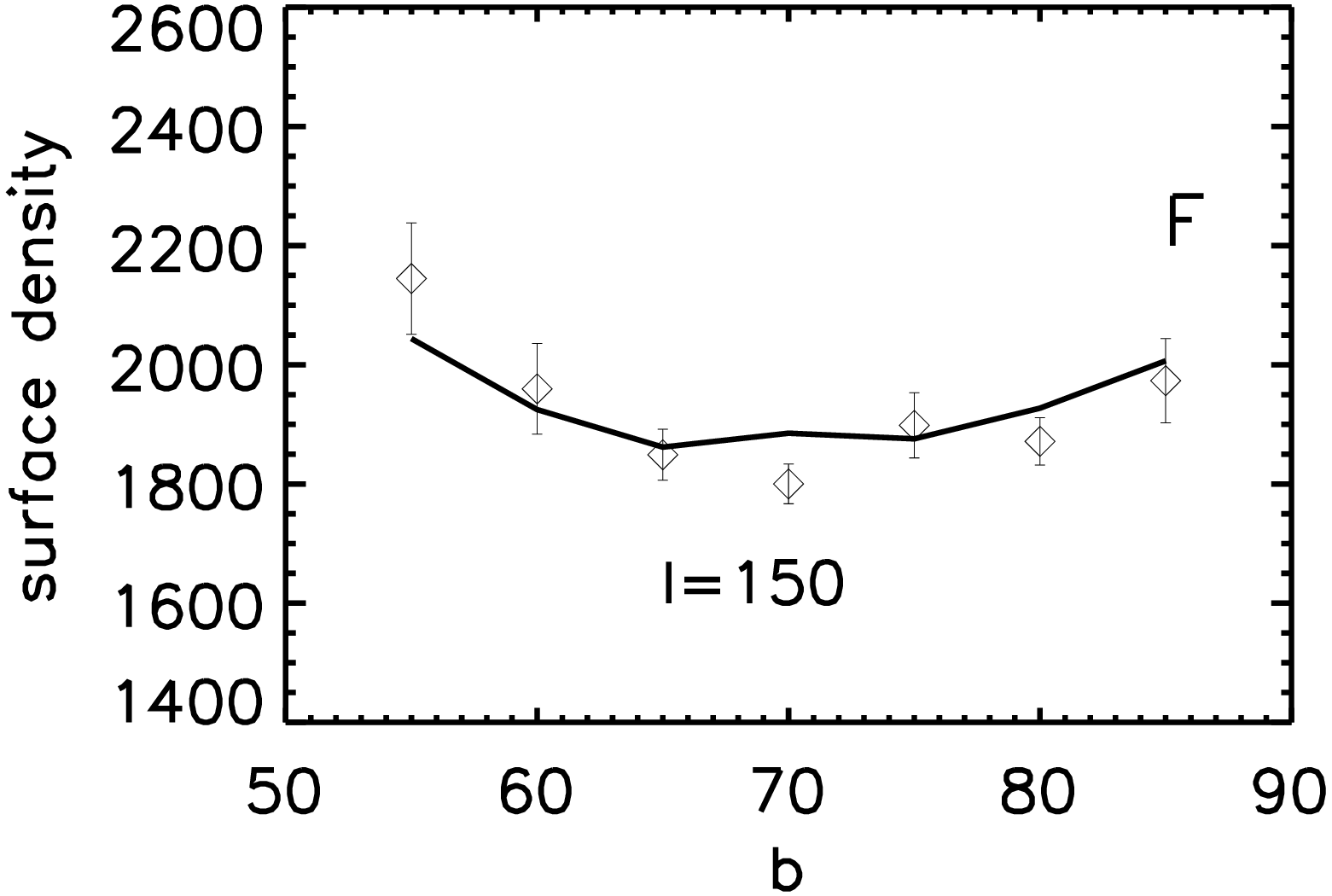}
\vspace{20pt}
 \caption{Fits of the surface number density using
groups 2--12. The solid and dashed lines are the theoretical
predictions, while the diamonds and triangles show the observational
data. Panel A: solid line, diamond, l=$0^{\circ}$; dashed line,
triangle, l=$180^{\circ}$. Panel B: solid line, diamond,
l=$30^{\circ}$; dashed line, triangle, l=$330^{\circ}$. Panel
C: solid line, diamond, l=$60^{\circ}$; dashed line, triangle,
l=$300^{\circ}$. Panel D: solid line, diamond, l=$120^{\circ}$;
dashed line, triangle, l=$240^{\circ}$. Panel E: solid line, diamond,
l=$90^{\circ}$; dashed line, triangle, l=$270^{\circ}$. Panel
F: solid line, diamond, l=$150^{\circ}$. }
\end{figure*}

\begin{figure*}
  \includegraphics[scale=0.35]{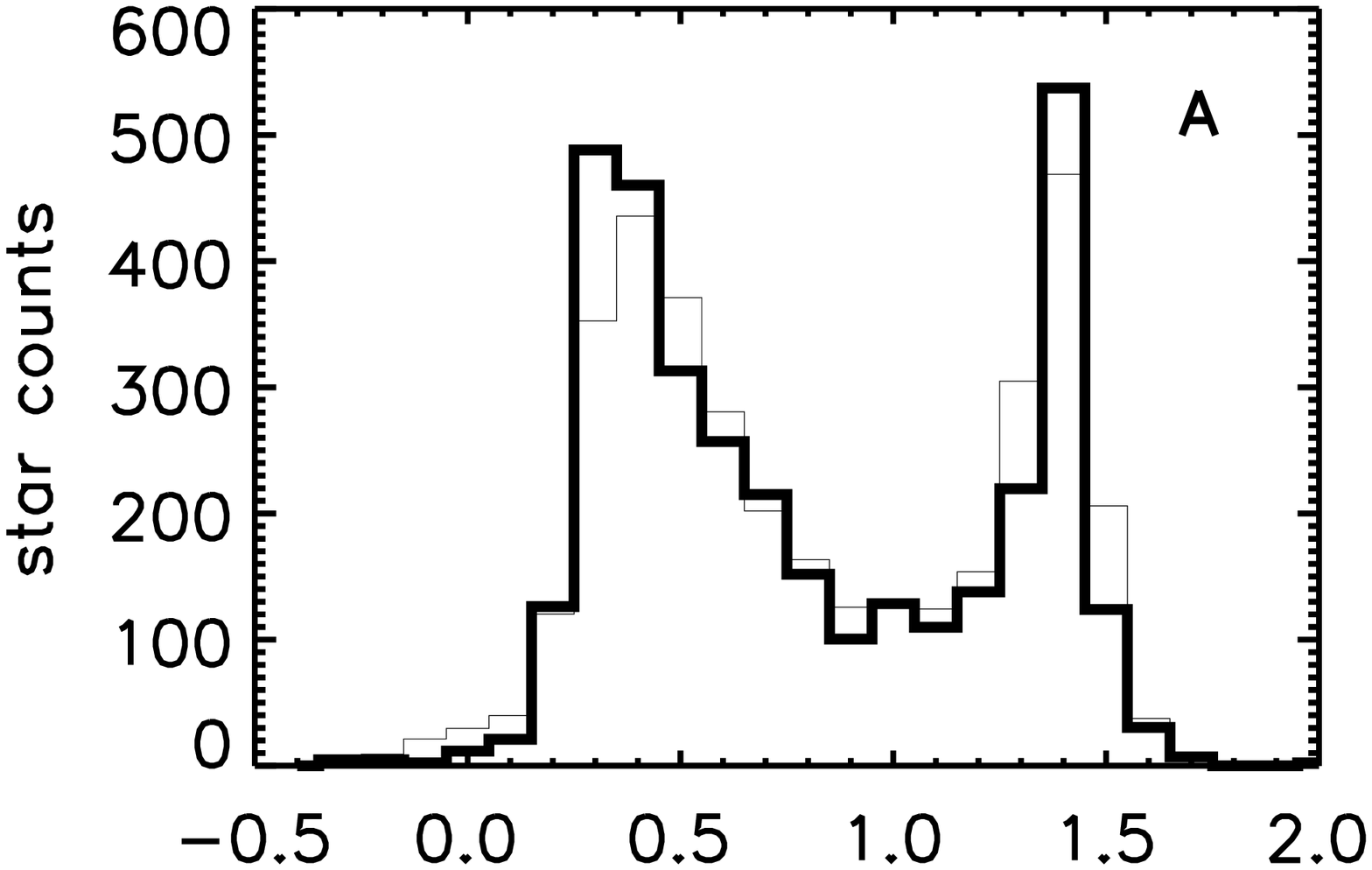}
  \hfill
  \includegraphics[scale=0.35]{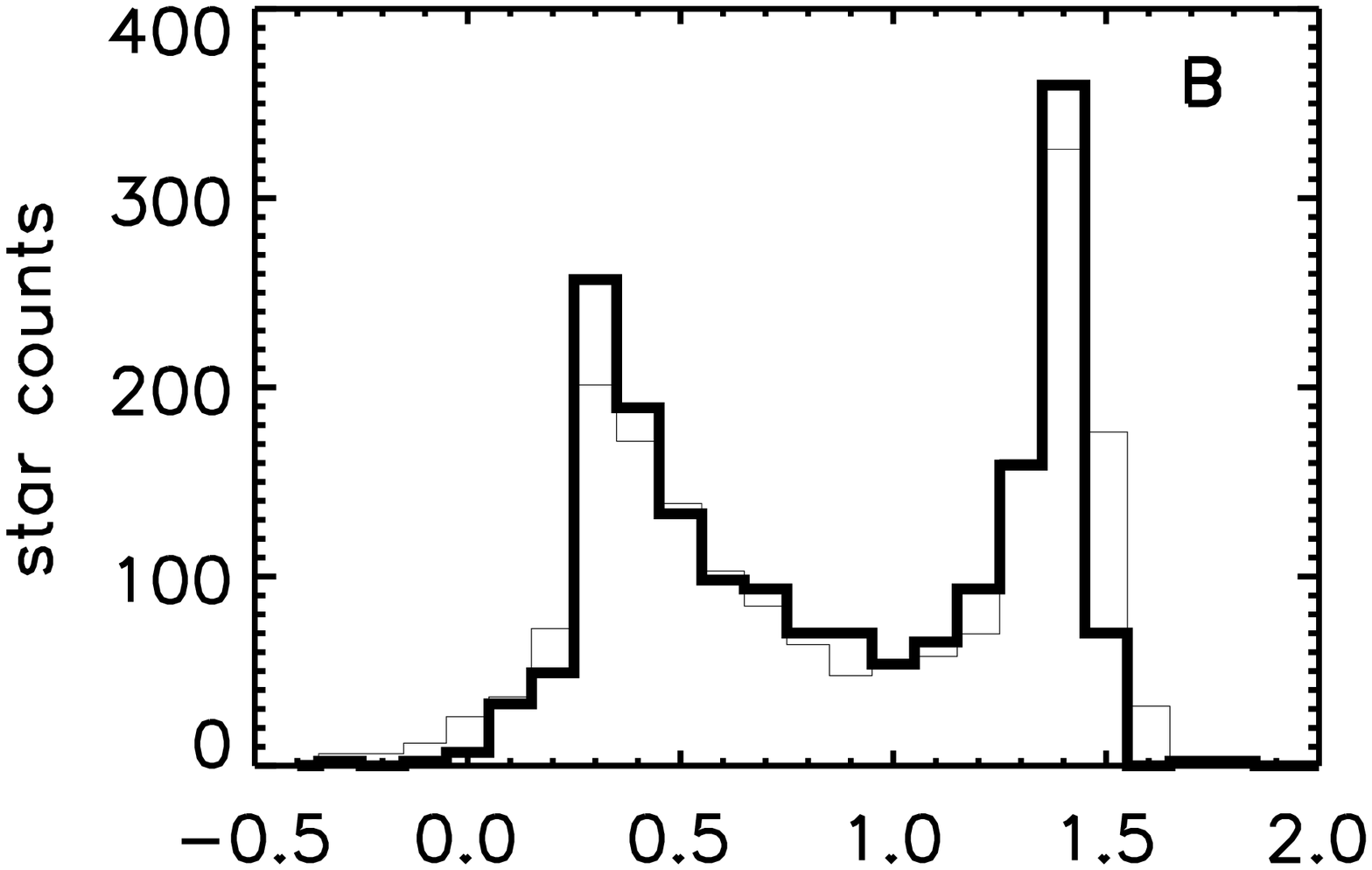}
  \vfill
  \includegraphics[scale=0.35]{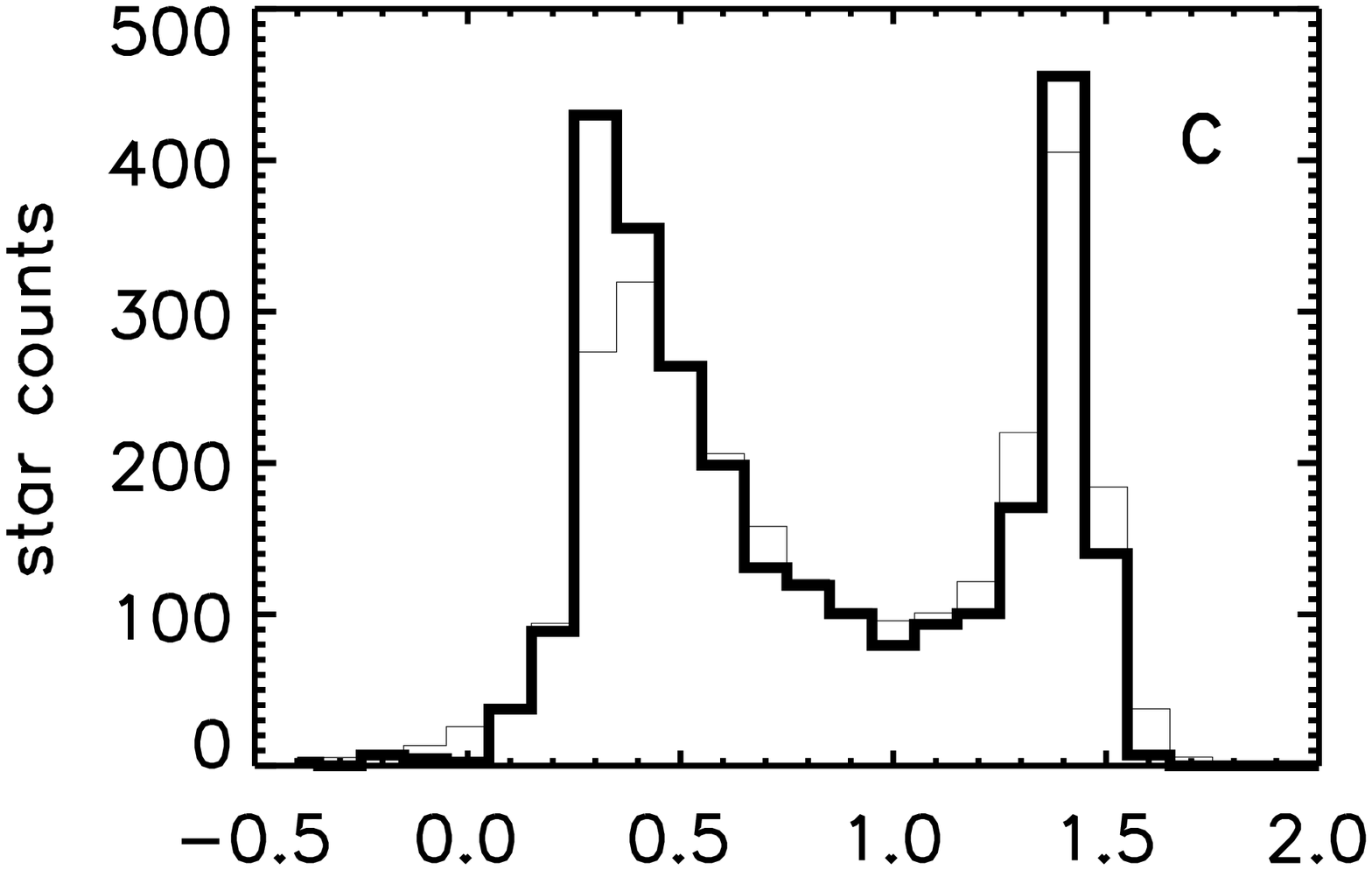}
  \hfill
  \includegraphics[scale=0.35]{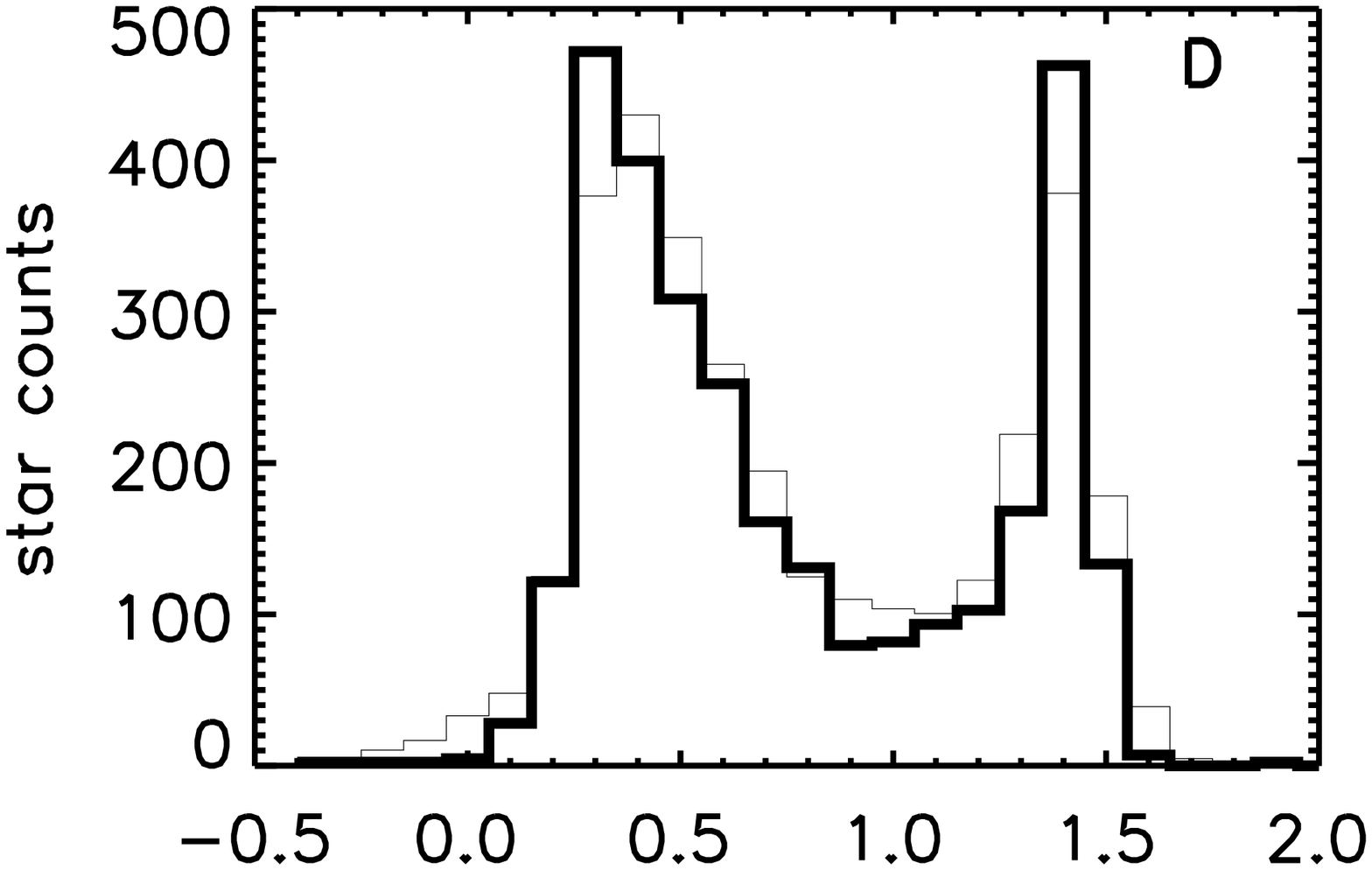}
  \vfill
  \includegraphics[scale=0.35]{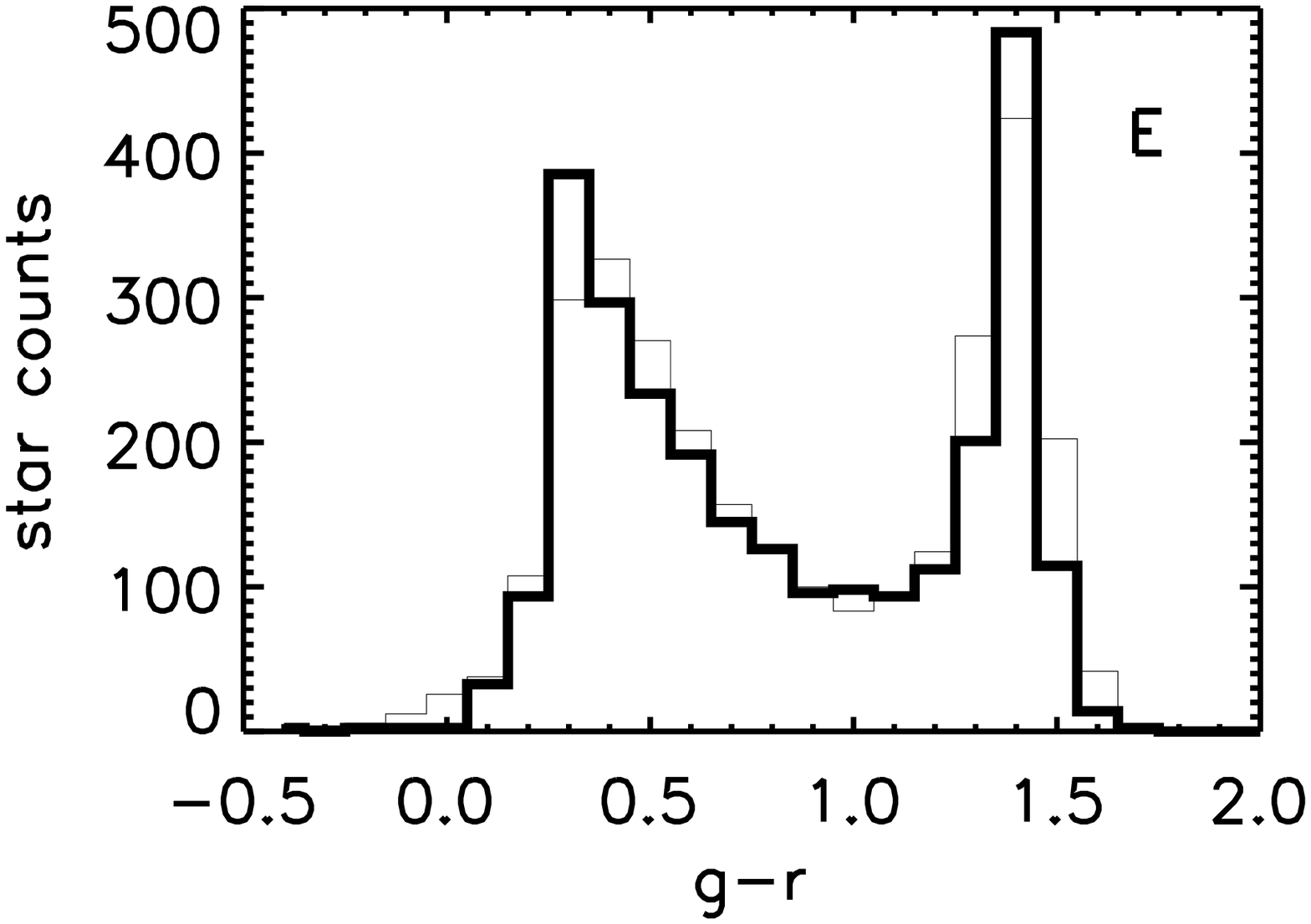}
  \hfill
  \includegraphics[scale=0.35]{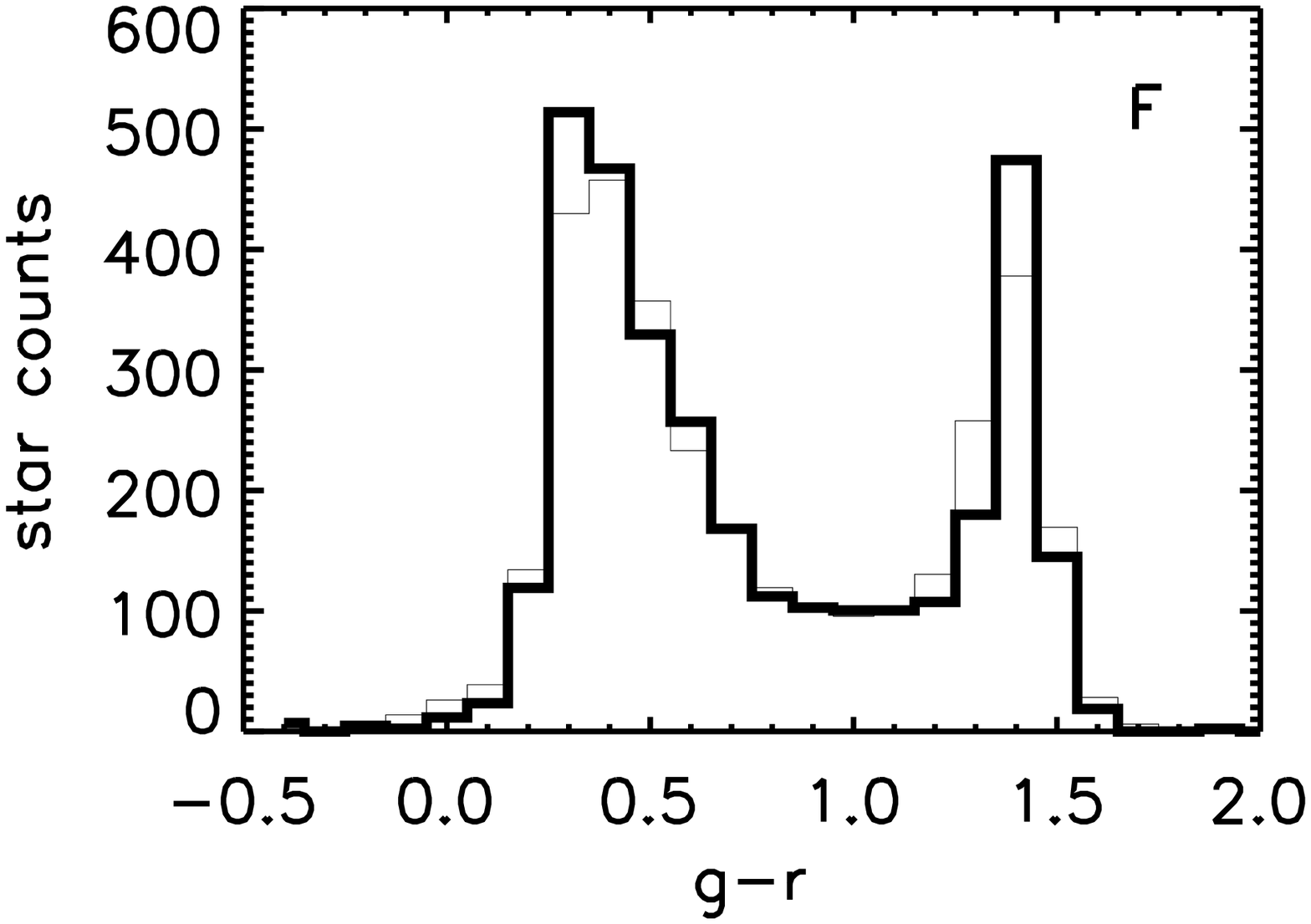}
   \vspace{20pt}
\caption{Fits of star count statistics in $(g-r)$ colour, where $g$ and $r$
are both within [22, 15] mag. The thin lines describe the
observational data, the solid black lines show the theoretical results. Panel
A: the results for ($0^{\circ},60^{\circ}$), Panel B:
($180^{\circ},60^{\circ}$), Panel C: ($30^{\circ},65^{\circ}$)
Panel D: ($330^{\circ},65^{\circ}$) Panel E:
($60^{\circ},60^{\circ}$) Panel F: ($300^{\circ},60^{\circ}$).}
\end{figure*}

\begin{figure*}
  \includegraphics[scale=0.35]{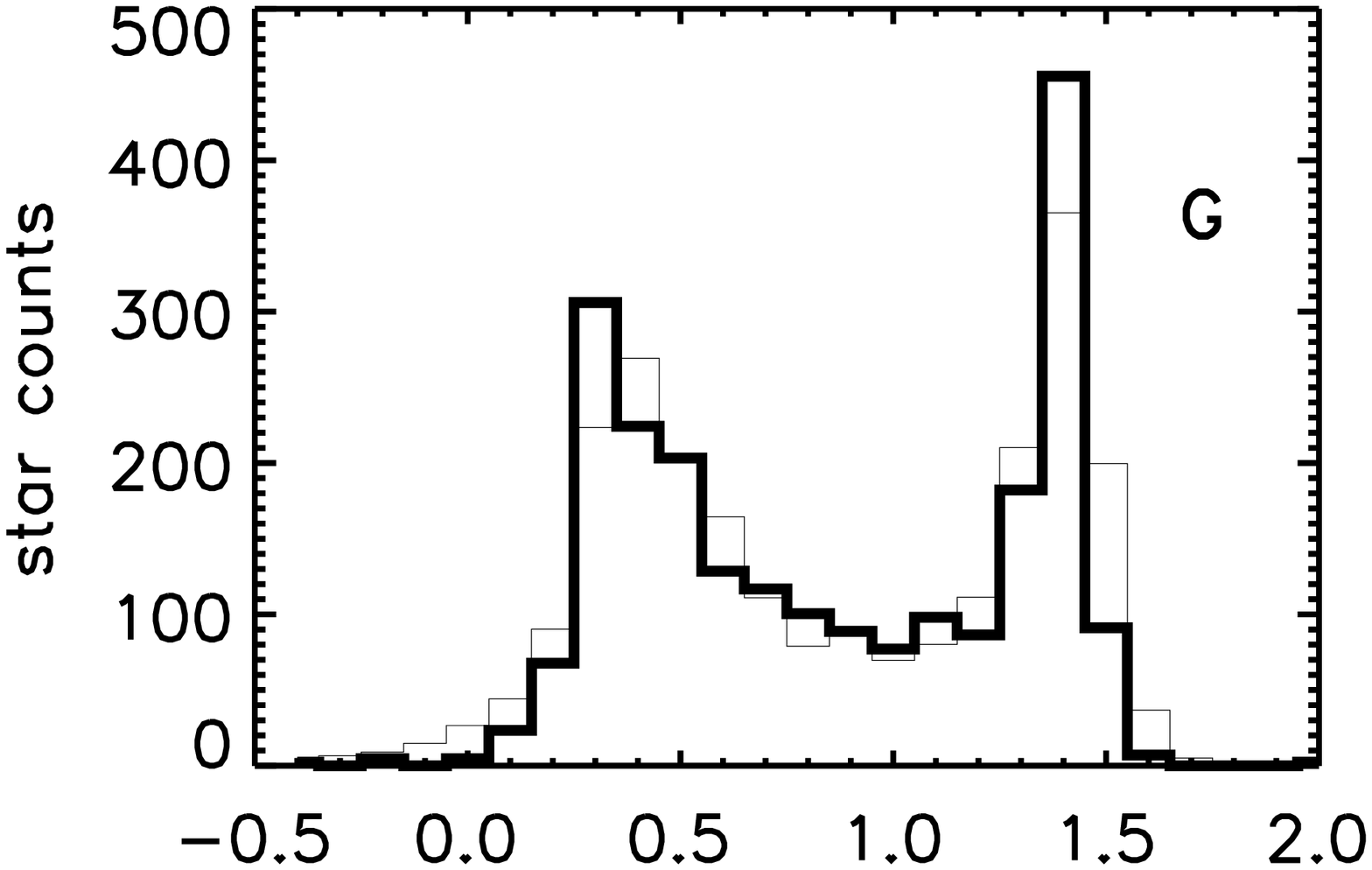}
  \hfill
  \includegraphics[scale=0.35]{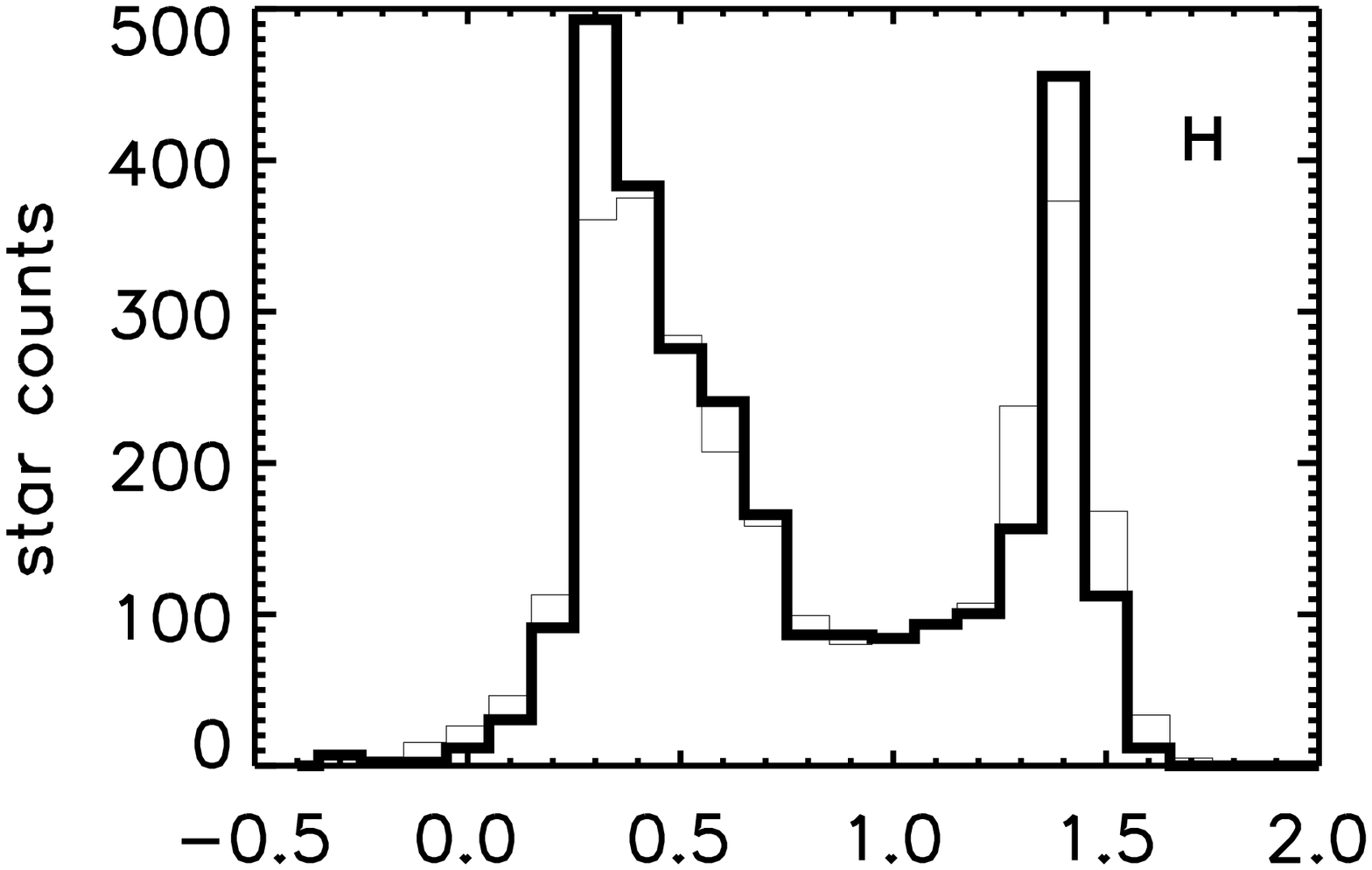}
  \vfill
  \includegraphics[scale=0.35]{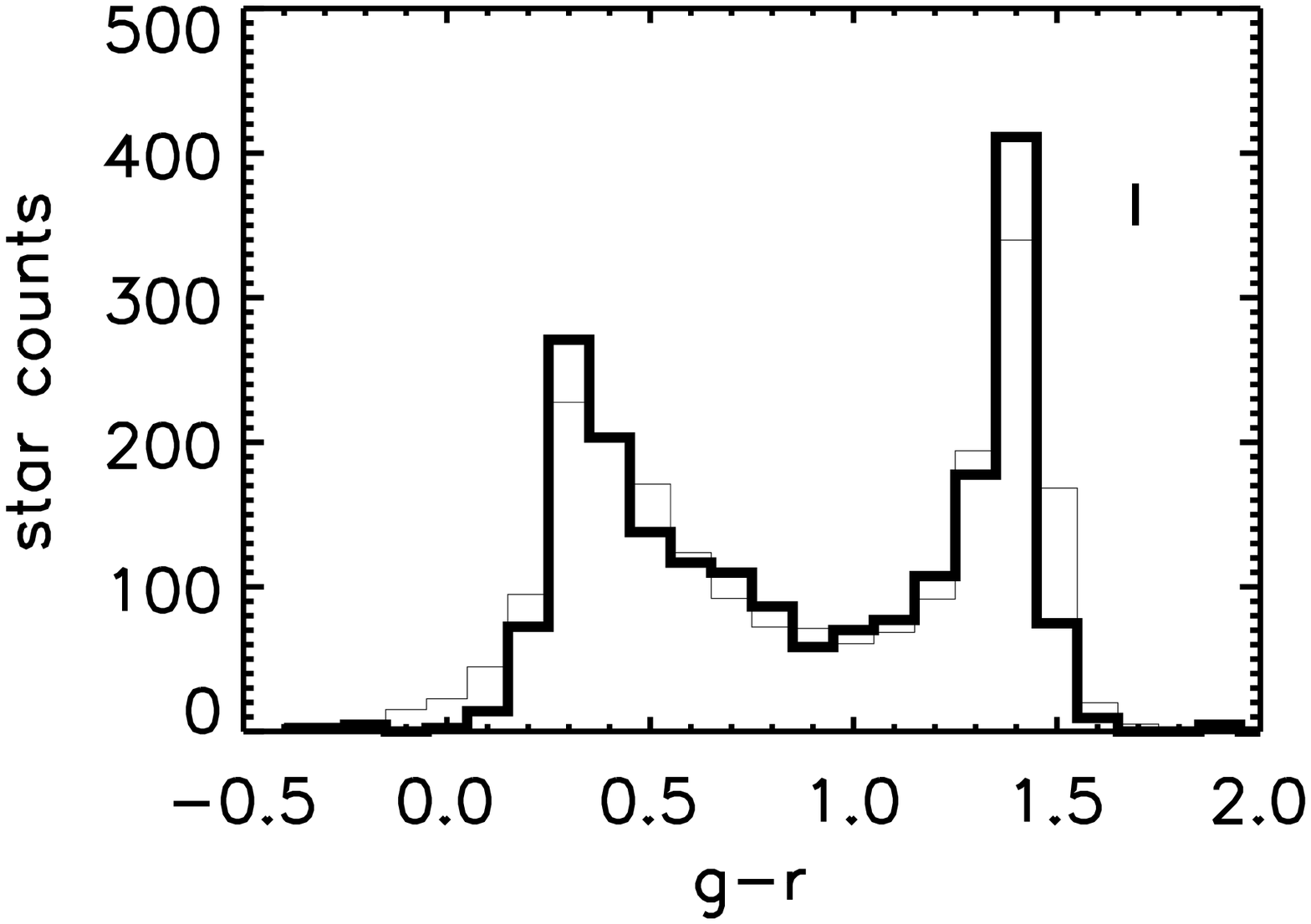}
  \hfill
  \includegraphics[scale=0.35]{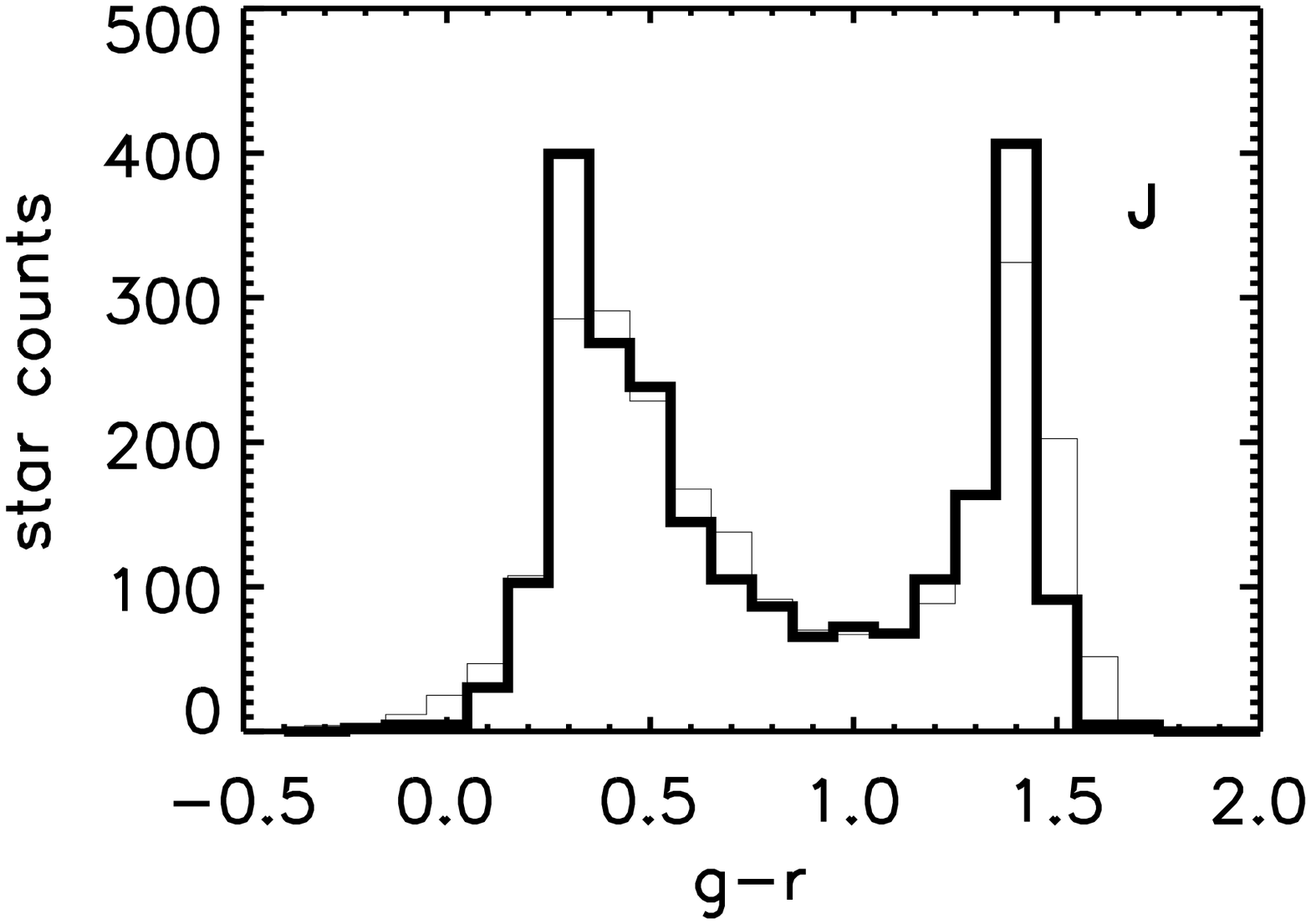}
  \vfill
  \includegraphics[scale=0.35]{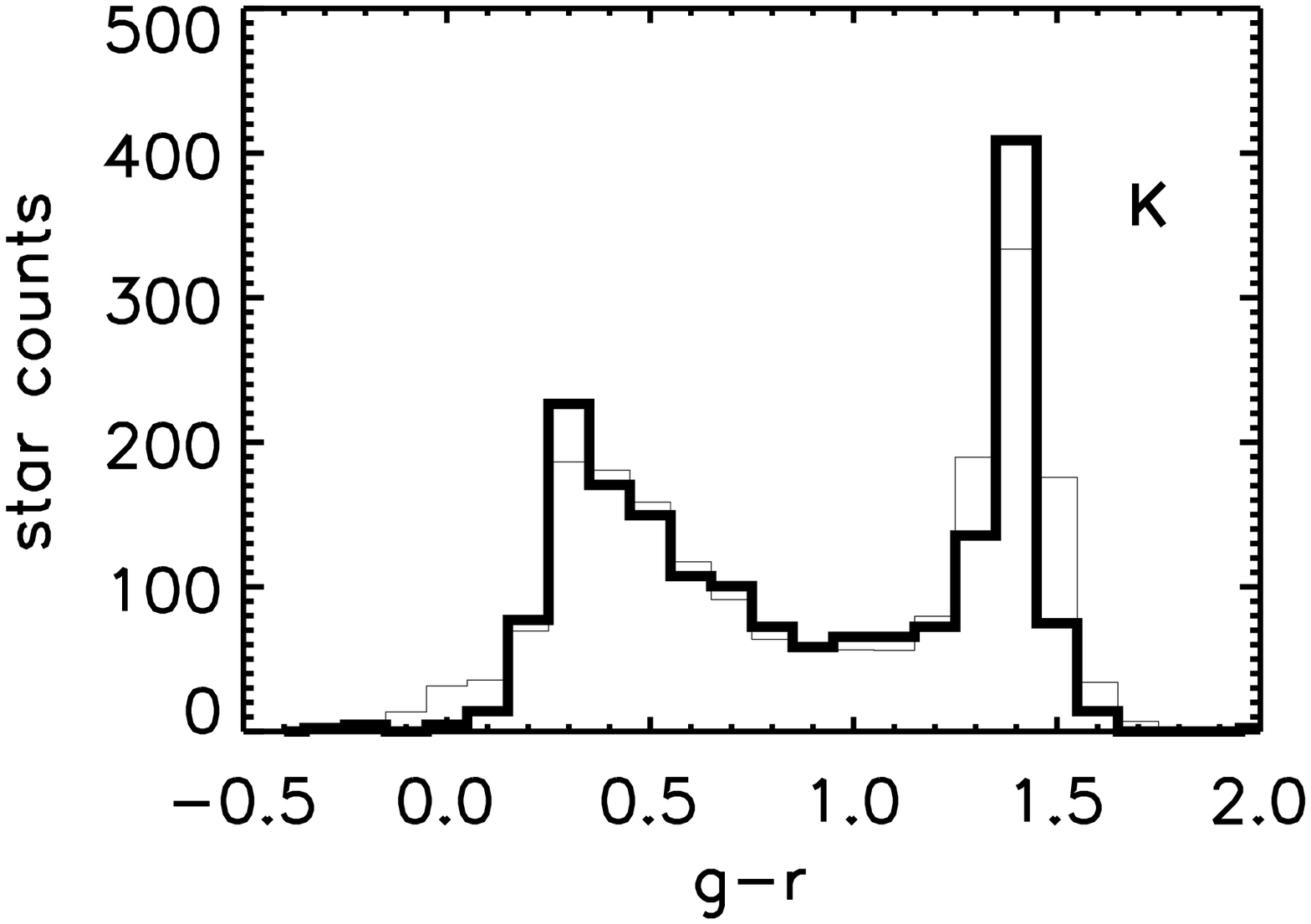}
\vspace{20pt}
 \caption{-continued. Panel G:
($90^{\circ},60^{\circ}$), Panel H: ($270^{\circ},60^{\circ}$),
Panel I: ($120^{\circ},60^{\circ}$) Panel
J:($240^{\circ},60^{\circ}$), Panel K: ($150^{\circ},60^{\circ}$)
}
\end{figure*}

\begin{table*}
 \centering
 \begin{minipage}{140mm}
\caption{The Galactic coordinates of the selected sky fields in
each group}
  \begin{tabular}{@{}ll@{}}
  \hline
 group & (l,b) \\
 \hline
 1 & ($0^\circ$,$60^\circ$),($10^\circ$,$60^\circ$),($40^\circ$,$60^\circ$),($50^\circ$,$60^\circ$),($60^\circ$,$60^\circ$),($70^\circ$,$60^\circ$),\\
   & ($80^\circ$,$60^\circ$),($90^\circ$,$60^\circ$),($100^\circ$,$60^\circ$),($110^\circ$,$60^\circ$),($120^\circ$,$60^\circ$),($130^\circ$,$60^\circ$),\\
   & ($140^\circ$,$60^\circ$),($150^\circ$,$60^\circ$),($160^\circ$,$60^\circ$),($170^\circ$,$60^\circ$),($180^\circ$,$60^\circ$),($190^\circ$,$60^\circ$),\\
   & ($200^\circ$,$60^\circ$),($230^\circ$,$60^\circ$),($240^\circ$,$60^\circ$),($250^\circ$,$60^\circ$),($260^\circ$,$60^\circ$),($270^\circ$,$60^\circ$),\\
   & ($280^\circ$,$60^\circ$),($290^\circ$,$60^\circ$),($300^\circ$,$60^\circ$),($310^\circ$,$60^\circ$),($320^\circ$,$60^\circ$),($330^\circ$,$60^\circ$),\\
   & ($340^\circ$,$60^\circ$),($350^\circ$,$60^\circ$)\\
 2 & ($0^\circ$,$55^\circ$), ($0^\circ$,$60^\circ$), ($0^\circ$,$65^\circ$), ($0^\circ$,$70^\circ$)\\
 3 & ($30^\circ$,$65^\circ$), ($30^\circ$,$70^\circ$), ($30^\circ$,$75^\circ$)\\
 4 & ($60^\circ$,$55^\circ$), ($60^\circ$,$60^\circ$), ($60^\circ$,$65^\circ$), ($60^\circ$,$70^\circ$), ($60^\circ$,$75^\circ$), ($60^\circ$,$80^\circ$), ($60^\circ$,$85^\circ$)\\
 5 & ($90^\circ$,$55^\circ$), ($90^\circ$,$60^\circ$), ($90^\circ$,$65^\circ$), ($90^\circ$,$70^\circ$), ($90^\circ$,$75^\circ$),($90^\circ$,$80^\circ$),($90^\circ$,$85^\circ$)\\
 6 & ($120^\circ$,$55^\circ$), ($120^\circ$,$60^\circ$), ($120^\circ$,$65^\circ$), ($120^\circ$,$70^\circ$), ($120^\circ$,$75^\circ$),($120^\circ$,$80^\circ$),($120^\circ$,$85^\circ$)\\
 7 & ($150^\circ$,$55^\circ$), ($150^\circ$,$60^\circ$), ($150^\circ$,$65^\circ$), ($150^\circ$,$70^\circ$),($150^\circ$,$75^\circ$), ($150^\circ$,$80^\circ$), ($150^\circ$,$85^\circ$)\\
 8 & ($180^\circ$,$55^\circ$), ($180^\circ$,$60^\circ$), ($180^\circ$,$65^\circ$), ($180^\circ$,$70^\circ$)\\
 9 & ($240^\circ$,$55^\circ$), ($240^\circ$,$60^\circ$), ($240^\circ$,$65^\circ$)\\
 10 & ($270^\circ$,$55^\circ$), ($270^\circ$,$60^\circ$), ($270^\circ$,$65^\circ$), ($270^\circ$,$70^\circ$), ($270^\circ$,$75^\circ$)\\
 11 & ($300^\circ$,$60^\circ$), ($300^\circ$,$65^\circ$), ($300^\circ$,$70^\circ$), ($300^\circ$,$75^\circ$)\\
 12 & ($330^\circ$,$60^\circ$), ($330^\circ$,$65^\circ$), ($330^\circ$,$70^\circ$), ($330^\circ$,$75^\circ$)\\
\hline
\end{tabular}
\end{minipage}
\end{table*}

\begin{table*}
 \centering
 \begin{minipage}{140mm}
\caption{The relative deviations of the sky field pairs for
$g,r\in[15,22]$ mag. (see text for details).}\label{tab2}
\begin{tabular}{@{}rrr@{}}
  \hline
 $\ell_1$ (degree)  & $\ell_2$ (degree)  & asymmetry ratio($\%$)   \\
  \hline
      10.000 &      350.000 &    7.49773 \\
      40.000 &      320.000 &     21.3195 \\
      50.000 &      310.000 &     19.9415 \\
      60.000 &      300.000 &     11.2650 \\
      70.000 &      290.000 &     20.6879 \\
      80.000 &      280.000 &     23.9994 \\
      90.000 &      270.000 &     18.9822 \\
      100.000 &      260.000 &     14.3713 \\
      110.000 &      250.000 &    9.98441 \\
      120.000 &      240.000 &     16.5926 \\
      130.000 &      230.000 &     14.8426 \\
      160.000 &      200.000 &     13.5624 \\
      170.000 &      190.000 &    5.12844 \\
\hline
\end{tabular}
\end{minipage}
\end{table*}

\begin{table*}
 \centering
 \begin{minipage}{140mm}
\caption{The relative deviations of the sky field pairs, similar to
Table~\ref{tab2} but for $g,r\in[19,21]$ mag.}\label{tab3}
\begin{tabular}{@{}rrr@{}}
  \hline
 $\ell_1$ (degree)  & $\ell_2$ (degree)  & asymmetry ratio($\%$)   \\
  \hline
      10.000 &      350.000 &     10.2431 \\
      40.000 &      320.000 &     32.7824 \\
      50.000 &      310.000 &     26.8430 \\
      60.000 &      300.000 &     24.9343 \\
      70.000 &      290.000 &     35.5734 \\
      80.000 &      280.000 &     33.3278 \\
      90.000 &      270.000 &     29.9915 \\
      100.000 &      260.000 &     26.4650 \\
      110.000 &      250.000 &     15.3487 \\
      120.000 &      240.000 &     16.3516 \\
      130.000 &      230.000 &     12.8257 \\
      160.000 &      200.000 &     19.2948 \\
      170.000 &      190.000 &    5.18586 \\
\hline
\end{tabular}
\end{minipage}
\end{table*}

\begin{table*}
 \centering
 \begin{minipage}{140mm}
\caption{Input ranges of the parameters.}
\begin{tabular}{@{}rrrr@{}}
  \hline
 parameter  & lower limit  & upper limit & step    \\
  \hline
n & 2 & 3 & 0.1 \\
p & 0.5 & 0.9 & 0.1 \\
q & 0.5 & 0.9 & 0.1\\
$\theta$ & $55^{\circ}$ & $65^{\circ}$ & $5^{\circ}$ \\
$\xi$ & $-20^{\circ}$ & $5^{\circ}$ & $5^{\circ}$ \\
$\phi$ & $-20^{\circ}$ & $5^{\circ}$ & $5^{\circ}$ \\
\hline
\end{tabular}
\end{minipage}
\end{table*}

\begin{table*}
 \centering
 \begin{minipage}{140mm}
\caption{The best-fit parameters for fixed power-law index $n$ at [2,2.6]}
\begin{tabular}{@{}rrrrrrrr@{}}
  \hline
 power-law index  & $\theta$ (degree)  & $p$ & $q$ & $\xi$ (degree) &
$\phi$ (degree) & $\chi^2$ & $\overline{\chi^2_{(g-r)}}$ \\
  \hline
  2.  & 65. & 0.5 & 0.5 & -15. &-5. & 2.302 & 7.092\\
  2.1 & 65. & 0.5 & 0.5 & -20. &-10. & 1.735 & 6.842\\
  2.2 & 60. & 0.5 & 0.5 & -10.& -10.&  2.02 & 6.632\\
  2.3 & 55. & 0.5 & 0.5 & 5.& -15. &  2.869 & 6.909\\
  2.4 & 65. & 0.6 & 0.5 & 5.& -20. &  2.965 & 7.073 \\
  2.5 & 65. & 0.7 & 0.6 & -15.& -10.& 3.055 & 7.204\\
  2.6 & 65. & 0.7 & 0.6 & -15.& -10.& 3.539 & 7.257\\
\hline
\end{tabular}
\end{minipage}
\end{table*}
\label{lastpage}


\begin{thebibliography}{}
\bibitem[\protect\citeauthoryear{Bahcall} {1980}]{b1}
        Bahcall, J. N., Soneira, R. M., 1980, ApJS, 44,73
\bibitem[\protect\citeauthoryear{Blitz}{1991}]{b2}
        Blitz, L., Spergel, D. N., 1991, ApJ, 370, 205
\bibitem[\protect\citeauthoryear{Chen} {2001}]{b3}
        Chen, B., Stoughton, C., Allyn Smith, J., et. al., 2001, ApJ, 553, 184
\bibitem[\protect\citeauthoryear{Crawford} {1975}]{b4}
        Crawford, D. L., 1975, AJ, 80, 955
\bibitem[\protect\citeauthoryear{Crawford} {1979}]{b5}
        Crawford, D. L., 1979, AJ, 84, 1858
\bibitem[\protect\citeauthoryear{Deutschman} {1976}]{b6}
        Deutschman, W. A., Davis, R. J., Schild, R. E., 1976, ApJS, 30, 97
\bibitem[\protect\citeauthoryear{Drimmel} {2003}]{b7}
        Drimmel, R., Cabrera-Lavers, A., L$\acute{o}$pez-Corredoira, M. , 2003, A\&A, 409, 205
\bibitem[\protect\citeauthoryear{Fukugita} {1996}]{b9}
        Fukugita, M., Ichikawa, T., Gunn, J. E., et. al., 1996, AJ, 111, 1748
\bibitem[\protect\citeauthoryear{Gilmore} {1984}]{b10}
        Gilmore, G., 1984, MNRAS, 207, 223
\bibitem[\protect\citeauthoryear{Girardi} {2004}]{b11}
        Girardi, L., Grebel, E. K., Odenkirchen, M., et. al., 2004, A\&A, 422, 205
\bibitem[\protect\citeauthoryear{Gould} {1996}]{b13}
        Gould, A., Bahcall, J. N., Flynn, C., 1996, ApJ, 465, 759
\bibitem[\protect\citeauthoryear{Hartwick} {2000}]{b14}
        Hartwick, F. D. A., 2000, AJ, 119, 2248
\bibitem[\protect\citeauthoryear{Hayes} {1978}]{b15}
        Hayes, D. S., 1978, proc. IAU Symp., 80, The HR diagram: The 100th anniversary of Henry Norris Russell, p. 65
\bibitem[\protect\citeauthoryear{Houk} {1975}]{b16}
        Houk, N., Cowley, A. P., 1975, Michigan Catalogue of Two-Dimensional Spectral Types for the HD stars Vol. 1, Univ Michigan, Ann Arbor
\bibitem[\protect\citeauthoryear{Houk} {1978}]{b17}
        Houk, N., Cowley, A. P., 1978, Michigan Catalogue of Two-Dimensional Spectral Types for the HD stars Vol. 2, Univ Michigan, Ann Arbor
\bibitem[\protect\citeauthoryear{Houk} {1982}]{b18}
        Houk, N., Cowley, A. P., 1982, Michigan Catalogue of Two-Dimensional Spectral Types for the HD stars Vol. 3, Univ Michigan, Ann Arbor
\bibitem[\protect\citeauthoryear{Jing} {2002}]{b19}
         Jing, Y. P., Suto, Y., 2002, ApJ, 574, 538
\bibitem[\protect\citeauthoryear{Juri$\acute{c}$}{2005}]{b20}
         Juri$\acute{c}$, M., Ivezi$\acute{c}$, $\check{Z}$., Brooks, A., et al., preprint (astro-ph/0510520)
\bibitem[\protect\citeauthoryear{Kron}{1978}]{b21}
        Kron, R. G., 1978, PhD thesis, University of California, Berkeley
\bibitem[\protect\citeauthoryear{Larsen}{1996}]{b22}
        Larsen, J. A., Humphreys, R. M., 1996, ApJ, 468, 99
\bibitem[\protect\citeauthoryear{Majewski}{1999}]{b23}
        Majewski, S. R., Siegel, M. H., Kunkel, W. E., et. al., 1999, AJ, 118, 1709
\bibitem[\protect\citeauthoryear{Mazzei} {2001}]{b24}
        Mazzei, P., Curir A., 2001, A\&A 372, 803
\bibitem[\protect\citeauthoryear{Newberg} {2002}]{b26}
         Newberg, H. J., Yanny, B., Rockosi, C., et. al. , 2002, ApJ, 569, 245
\bibitem[\protect\citeauthoryear{Newberg} {2005a}]{b27}
        Newberg, H. J., Yanny, B., preprint (astro-ph/0502386)
\bibitem[\protect\citeauthoryear{Newberg} {2005b}]{b28}
        Newberg, H. J., Yanny, B., preprint (astro-ph/0507671)
\bibitem[\protect\citeauthoryear{Parker} {2003}] {b29}
        Parker, J. E., Humphreys, R. M., Larsen, J. A., 2003, AJ, 126, 1346
\bibitem[\protect\citeauthoryear{Parker}{2004}] {b30}
        Parker, J. E., Humphreys, R. M., Beers, T. C., 2004, AJ, 127, 1567
\bibitem[\protect\citeauthoryear{Perterson} {1979}]{b31}
         Peterson, B. A., Ellis, R. S., Kibblewhite, E. J., et. al. , 1979, ApJ, 233, L109
\bibitem[\protect\citeauthoryear{Press} {1992}]{b33}
        Press, W. H., Teukolsky, S. A., Vetterling, W. T., et. al., 1992, Numerical recipes in Fortran: the art of scientific computing, Cambridge University Press, p. 616
\bibitem[\protect\citeauthoryear{Reid} {1993}]{b34}
        Reid, N., 1993, ASPC, 49, 37
\bibitem[\protect\citeauthoryear{Reid} {1996}]{b35}
        Reid, I. N., Yan, L., Majewski, S., et. al., 1996, AJ, 112, 1472
\bibitem[\protect\citeauthoryear{Reid}{2002}]{b36}
        Reid, I. N., 2002, AJ, 124, 2721
\bibitem[\protect\citeauthoryear{Robin} {1986}]{b37}
        Robin, A., Cr\'ez\'e, M., 1986, A\&A, 157,71
\bibitem[\protect\citeauthoryear{Seares} {1925}]{b38}
         Seares, F. H., 1925, ApJ, 61, 114
\bibitem[\protect\citeauthoryear{Siegel} {2002}]{b39}
        Siegel, M. H., Majewski, S. R., Reid, I. N., et. al., 2002, ApJ, 578, 151
\bibitem[\protect\citeauthoryear{Spergel}{1988}]{b40}
        Spergel, D. N.,  Blitz, L., 1988, BAAS, 20R1017S
\bibitem[\protect\citeauthoryear{Stoughton}{2002}]{b41}
        Stoughton, C., Lupton, R. H., Bernardi, M., et al., 2002, AJ, 123, 485
\bibitem[\protect\citeauthoryear{Tyson} {1979}]{b42}
        Tyson, J. A., Jarvis, J. F., 1979, ApJ, 230, 153
\bibitem[\protect\citeauthoryear{Wielen} {1974}]{b44}
        Wielen, R., 1974, Highlights of Astronomy, 3, p. 395, ed. G. Contopoulos
\bibitem[\protect\citeauthoryear{Wielen} {1983}]{b45}
        Wielen, R., Jahreiss, H., kruger, R., 1983, in Davis Phillip, A. G., Upgren, A. R., eds, Proc. IAU Colloq. 76, Nearby Stars and the Stellar Luminosity Function, p. 163
\bibitem[\protect\citeauthoryear{Wyse} {2005}]{b46}
        Wyse, R. F. G., Gilmore, G., preprint (astro-ph/0510025)
\bibitem[\protect\citeauthoryear{Zheng} {2001}]{b47}
        Zheng, Z., Flynn, C., Gould, A., et al., 2001, ApJ, 555, 393
\bibitem[\protect\citeauthoryear{Zheng} {2004}]{b48}
        Zheng, Z., Flynn, C., Gould, A., et. al., 2004, ApJ, 601, 500

\end{thebibliography}
\end{document}